\documentclass[aps,prd,twocolumn,showpacs,nofootinbib,amsmath,amssymb,amsfonts,showkeys]{revtex4-2}

\usepackage{bm}
\usepackage{mathtext}
\usepackage[T1,T2A]{fontenc}
\usepackage[utf8]{inputenc}
\usepackage[english]{babel}
\usepackage[dvips]{graphicx}
\usepackage{adjustbox}

\usepackage{xcolor}

\begin{document}

\title{On the Possibility of Detecting a Global Signal in the Line \\ of the Hyperfine Structure of Hydrogen from the Dark Ages }

\author{О. Konovalenko$^1$, V. Zakharenko$^1$, B. Novosyadlyj$^{2,3}$, L.I. Gurvits$^{4,5}$, S. Stepkin$^1$, \\ Y. Vasylkivskyi$^1$, P. Tokarsky$^1$, О. Ulyanov$^1$, O. Stanislavsky$^1$, І. Bubnov$^1$}
\medskip

\affiliation{\it $^1$Institute of Radio Astronomy of NASU, 4 Mystetstv str., 61002 Kharkiv, Ukraine}
\affiliation{$^2$Astronomical Observatory of Ivan Franko National University of Lviv, \\
8 Kyryla I Methodia str., Lviv 79005, Ukraine }
\affiliation{\it $^3$College of Physics and International Center of Future Science of Jilin University, \\Qianjin Street 2699, Changchun 130012, P. R. China}
\affiliation{\it $^4$Joint Institute for VLBI ERIC, Oude Hoogeveensedijk 4, 7991 PD Dwingeloo, The Netherlands}
\affiliation{\it $^5$Faculty of Aerospace Engineering, Delft University of Technology, Kluyverweg 1, 2629 HS Delft, The Netherlands}

\date{\today} 

\begin{abstract}
We analyze the possibilities of detecting a signal in the hydrogen 21~cm line, which was formed in the early  Universe during the the Dark Ages cosmological epoch, using the Ukrainian radio telescopes UTR-2 and GURT of the National Academy of Sciences of Ukraine. As a result of cosmological expansion, this line is shifted to the decameter range of wavelengths ($\lambda_{obs}\approx18$ m, $\nu_{obs}\approx16$ MHz) and is in the band of operational frequencies of these telescopes. The brightness temperature of the predicted sky-averaged global signal ranges from $\sim-0.08$ to $\sim0.02$ K, depending on the cosmological model. Such a faint signal is a challenge even for the world's largest radio telescope in the decameter wavelength range, UTR-2, since the signal level of the foreground synchrotron radiation of the Galaxy at these wavelengths is 20000--40000~K. The paper highlights the peculiarities of spectroscopy at the decameter waves, interfering factors of natural and instrumental origin and the ways of eliminating them in order to reliably detect the signal in the 21~cm line, which can become an important source of information both about the environment in which the first stars and galaxies were born, and about the nature of dark matter particles and the magnitude of primordial magnetic fields. It was concluded that the detection of such a signal using the most sensitive radio telescopes at the decameter wavelength range is possible (with the signal integration over the frequency band of 25~MHz), the detection time will be $\sim50$~days) and can be implemented in the coming years of peace in Ukraine. 
\end{abstract}
\maketitle
 
\section*{Introduction} 

Cosmological research is one of the most challenging areas of modern astronomy. This is true for both theoretical and experimental research for relevant effects. The reasons for this encompasses the need to broaden the classical methods of theoretical physics and astrophysics, the uncertainty of initial conditions, the remoteness of the objects under study, and the faintness of effects in any range of cosmic radiation of all-wave astronomy.

The present paper suggests consolidating the efforts of theorists and  experimentalists aimed at assessing the possibilities of the existence and detection of a new cosmological phenomenon, i.e. the neutral hydrogen absorption lines (at the rest-frame 21~cm wavelength), which, due to the cosmological expansion, is redshifted to the range of decameter waves.

The research is based on the experience of the Lviv school of theoretical astrophysics and cosmology, as well as the achievements in the field of decameter radio astronomy, which has been developing in Kharkiv for more than half a century. Unfortunately, the world-famous S.\,Ya.~Braude Radio Astronomical Observatory was destroyed as a result of the aggression by the Russian Federation in 2022. Owing to the heroic actions of the Armed Forces of Ukraine, the observatory was liberated in September 2022. Now the infrastructure and research facilities of the Observatory are being restored gradually.

\section{21 cm Line from the Dark Ages}
 
Сosmic microwave background (CMB), experimentally discovered by Arno Penzias and Robert Wilson in 1965 \cite{Penzias1965}, is a key source of information about our Universe and the formation of its large-scale structure. The study of spatial fluctuations of its temperature and polarization by COBE\footnote{Cosmic Background Explorer (1989-1993), https://lambda.gsfc.nasa.gov/product/cobe/.}, WMAP\footnote{Wilkinson Microwave Anisotropy Probe (2001-2010), https://map.gsfc.nasa.gov/.}, and Planck\footnote{Space observatory Planck (2009-2018), https://sci.esa.int/web/planck/} space missions ushered in the era of precision cosmology, when the parameters of the cosmological model are determined to within a percent \cite{Planck2020a,Planck2020b}. At the same time, fundamental questions about the nature of the dominant components of the Universe in terms of average density -- dark energy and dark matter, the causes of baryon asymmetry, inflation models, as well as the formation of the first stars and galaxies -- remain unanswered. The latter problem has been exacerbated by the discovery of massive galaxies with intense star formation at redshifts $\sim13$ using the James Webb Space Telescope \cite{Robertson2023,Curtis-Lake2023}. On the other hand, the refinement of the cosmological distance scale revealed a discrepancy between the values of the Hubble’s constant according to CMB and type Ia supernovae at the level of  $\sim4\sigma$ \cite{Verde2019,Riess2022}. These discrepancies indicate the need to search for new cosmological tests, in particular, to analyze the possibilities of detecting the 21~cm hyperfine structure line of neutral hydrogen from the Dark Ages and the Cosmic Dawn cosmological epochs, which preceded the formation of the stars and galaxies we observe today. This line is an important channel of information about the state of baryonic matter (temperature, concentration, ionization, spatial inhomogeneities, etc.) in the epoch of the formation of the first stars and galaxies (see, for example, reviews \cite{Bromm2011,Pritchard2012}). In recent decades, experiments have been conducted to record the global (sky-averaged) signal in this line, which has been shifted by the cosmological expansion of the Universe at the meter--decameter wavelength range. In 2018, the first registration of this signal at a frequency of 78 MHz was announced in the EDGES experiment (Experiment to Detect the Global Epoch of Reionization Signature) \cite{Bowman2018}. However, recently published data from a similar experiment, SARAS~3 (Shaped Antenna measurement of the background RAdio Spectrum 3) \cite{Singh2022}, failed to confirm this discovery. 

For the past several years, scientists from the Astronomical Observatory of the Ivan Franko National University of Lviv and the Institute of Radio Astronomy of the National Academy of Sciences of Ukraine, jointly with peers from other institutions, have been conducting theoretical estimates of the intensity of radio emission from the Dark Ages and the Cosmic Dawn in the lines of the first molecules \cite{Novosyadlyj2020,Kulinich2020,Novosyadlyj2022} and hydrogen atoms \cite{Novosyadlyj2020a,Novosyadlyj2023}. In \cite{Novosyadlyj2020a}, the intensity of the luminosity of protogalaxies (halos) in the 21 cm line formed at redshifts $z\sim30-10$ was estimated. In \cite{Novosyadlyj2023}, the dependence of the global signal in the 21 cm line during the Dark Ages, Cosmic Dawn, and reionization on the parameters of cosmological models and the first light was studied. It is shown that the registration of the spectral features of the line at different wavelengths can be a good test for the parameters of cosmological models, the nature of dark matter particles, and the strength of the primary magnetic field.

In the accompanying paper \cite{Novosyadlyj2023a}, we showed that in the standard $\Lambda$CDM model with post-Planckian parameters an absorption line $\lambda_{HI}=21$~cm in the CMB spectrum is formed in the Dark Ages at the redshift $z=87$ and, for an Earth-based observer, is shifted to the decameter wavelength range: $\lambda_{obs}=\lambda_{HI}(1+z)\approx18$~m. The value of the differential brightness temperature in the center of the absorption line $\delta T_{br}\approx-35$ mK, the frequency at the absorption maximum is 16~MHz, and the effective line width is $\approx$25~MHz. The line depth is moderately sensitive to $\Omega_b$ and $H_0$, faintly sensitive to $\Omega_{dm}$ and insensitive to the other parameters of the standard $\Lambda$CDM model. However, the line is very sensitive to additional heating or cooling of baryonic matter during the Dark Ages, so it can be an effective test for non-standard cosmological models. In the models with decaying or self-annihilating dark matter, as well as with primordial global stochastic magnetic fields, the higher the density of these dark matter components and the magnetic field strength, the higher the temperature of baryonic matter during this period. The absorption line becomes shallower, disappears, and turns into emission line at values of the component parameters below the upper limits set for them, as follows from the available observational data.

The largest and the most sensitive radio telescope in this wavelength range is the Ukrainian T-shaped Radio telescope (UTR-2) and the Giant Ukrainian Radio Telescope (GURT), the development of which is being completed. In this paper, we analyze the possibilities of using them for detecting a global signal in the line of the hyperfine structure of the basic state of hydrogen from the Dark Ages, shifted to the decameter wavelength range.

As noted above, observational searches for such lines have been repeatedly conducted in the 50--120~MHz band, with successful detection at 78~MHz reported in the EDGES experiment but refuted in the SARAS~3 experiment. In recent years, tools and methods for such a search in the 20--200~MHz band have been actively developed, taking into account the extremely low intensity of the expected signals and the large number of interfering phenomena inherent in low-frequency radio astronomy \cite{Tauscher2018,Rapetti2020,Tauscher2020,Tauscher2021}. Among the main scientific objectives of many new generation low-frequency radio telescopes is also the search for relevant cosmological spectral features at lower frequencies. 

The use of pattern recognition methods or, in other terminology, ``machine learning'' and ``artificial intelligence'' is becoming increasingly popular. At the same time, the radio spectroscopic and interference features of decameter-wave radio astronomy at $<30$ MHz wavelengths, where the cosmological line is at $\sim16$~MHz \cite{Novosyadlyj2023a} and which is the proposed for search in this paper, are underrepresented in the studies and publications available to date.
On the other hand, 50~years of experience in the development of observational tools and methods of low-frequency radio astronomy, as well as the use of the above-mentioned Ukrainian radio telescopes, allow us to reassess the possibilities of detecting this interesting and important cosmological feature and to propose an adequate observational methodology.

\section{Radio astronomy at decameter waves in Ukraine}
 
It is well known that Ukraine is one of the global leaders in decameter radio astronomy ($\lambda = 10 - 100\,\, \mbox{m},\, \nu = 3-30\,\, \mbox{MHz}$). This
is universally recognized, has been maintained for about 60 years, and is based on the creation and use of the world's largest in this frequency domain radio telescope UTR-2 (with a record-breaking effective area of 150 000 square meters) and the construction of the unique Ukrainian Radio Interferometer of the Academy of Sciences (URAN) system on its basis \cite{1,2}. The founder of this field of science, which is now becoming increasingly important, was the prominent Ukrainian scientist S.~Ya.~Braude (1911--2003). We cannot help mentioning with gratitude another unique person, President of the National Academy of Sciences of Ukraine B.~Ye.~Paton (1918--2020), who until the last days of his bright life fully supported the development of low-frequency radio astronomy in Ukraine.

Over the past decades, UTR-2 and URAN have been upgraded continuously, including antenna systems, analog and digital equipment, as well as observation methods. The most advanced information and telecommunication technologies were used, which led to a thousandfold increase in the efficiency and meaningfulness of national radio astronomy \cite{3,4}. Moreover, next to UTR-2, a new-generation radio telescope GURT is being built, which has a much wider wavelength band of 8--80 MHz (this is a combined decameter-meter range) and offers many other advantages \cite{3,5}.

\begin{figure*}[htb]
\includegraphics[width=0.48\textwidth]{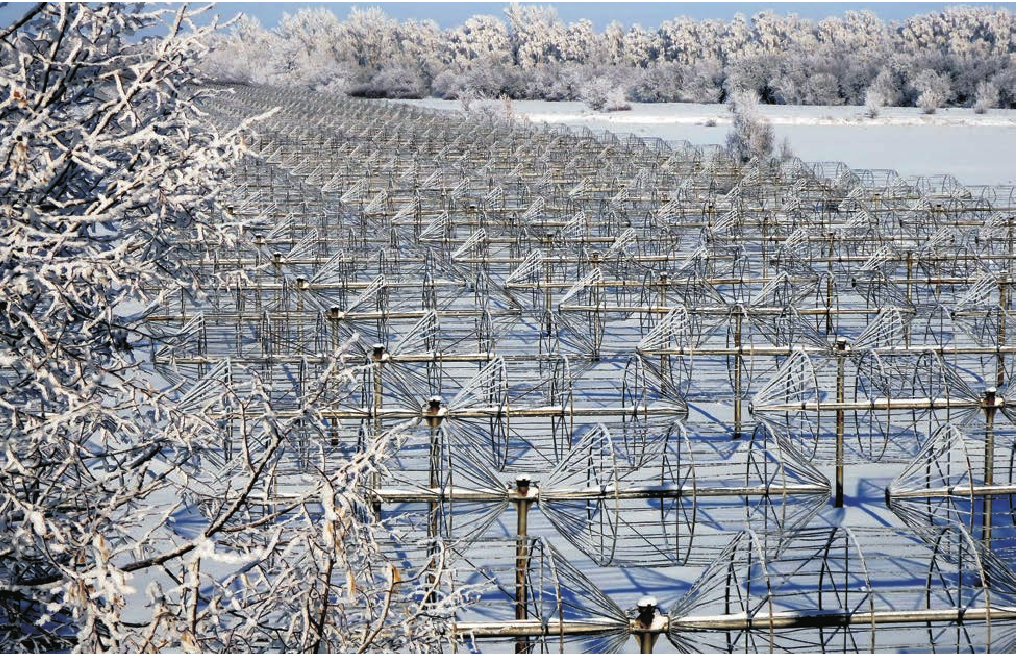} 
\includegraphics[width=0.465\textwidth]{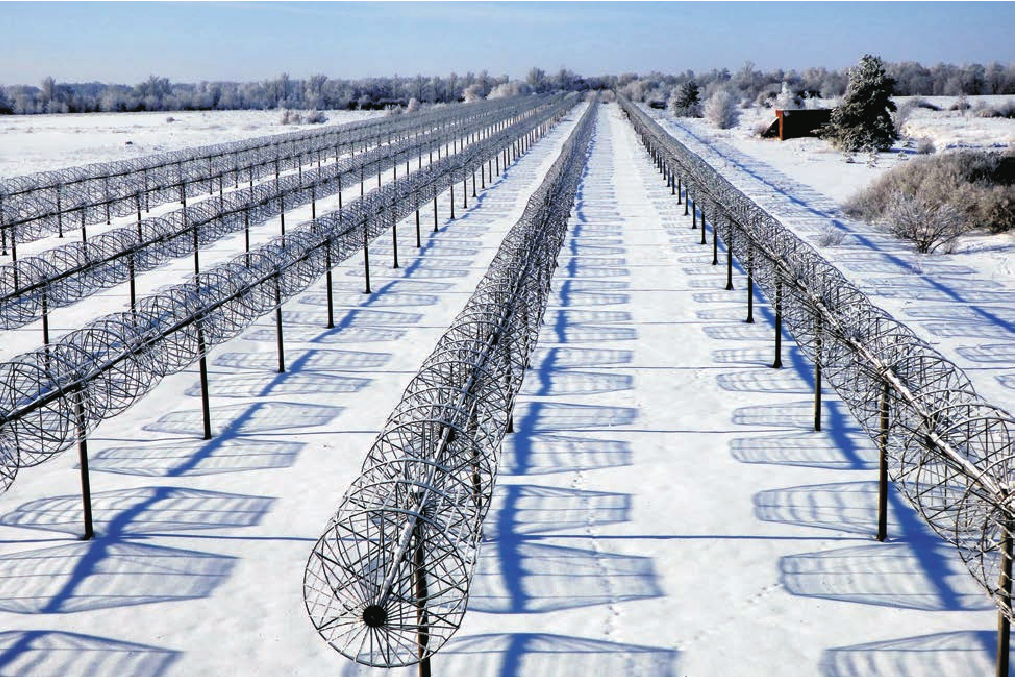}
\caption{The North-South (left) and East-West (right) antennas of the UTR-2 radio telescope (2020 photo)}
\label{rt1}
\end{figure*}

\begin{figure}[htb]
\includegraphics[width=0.48\textwidth]{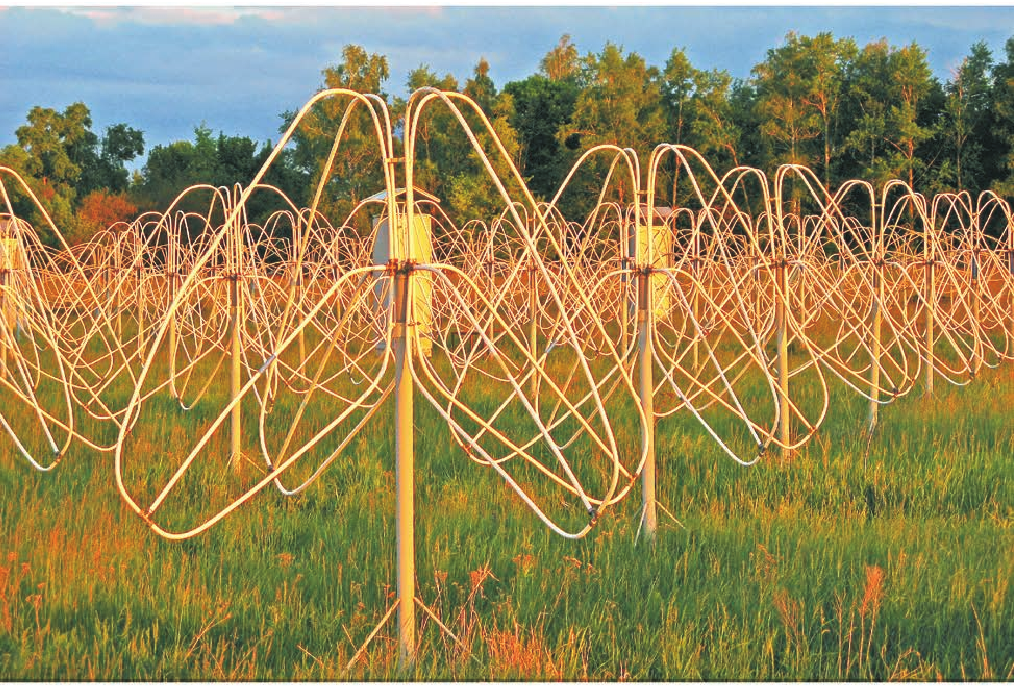}  
\caption{Wideband radio telescope of the new generation GURT.}
\label{rt2}
\end{figure}

\begin{figure}[htb]
\includegraphics[width=0.48\textwidth]{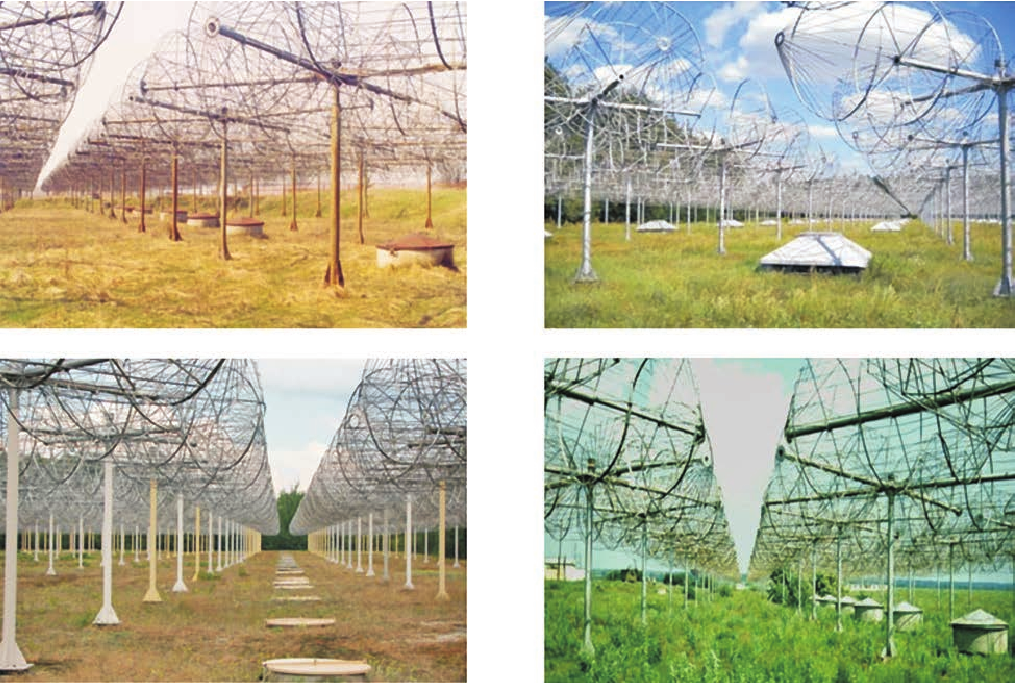}  
\caption{Radio telescopes URAN-1 (RI NASU), URAN-2 (PGO IGF NASU), URAN-3 (PMI NASU), URAN-4 (RI NASU).}
\label{rt3}
\end{figure}

\begin{figure}[htb]
\includegraphics[width=0.48\textwidth]{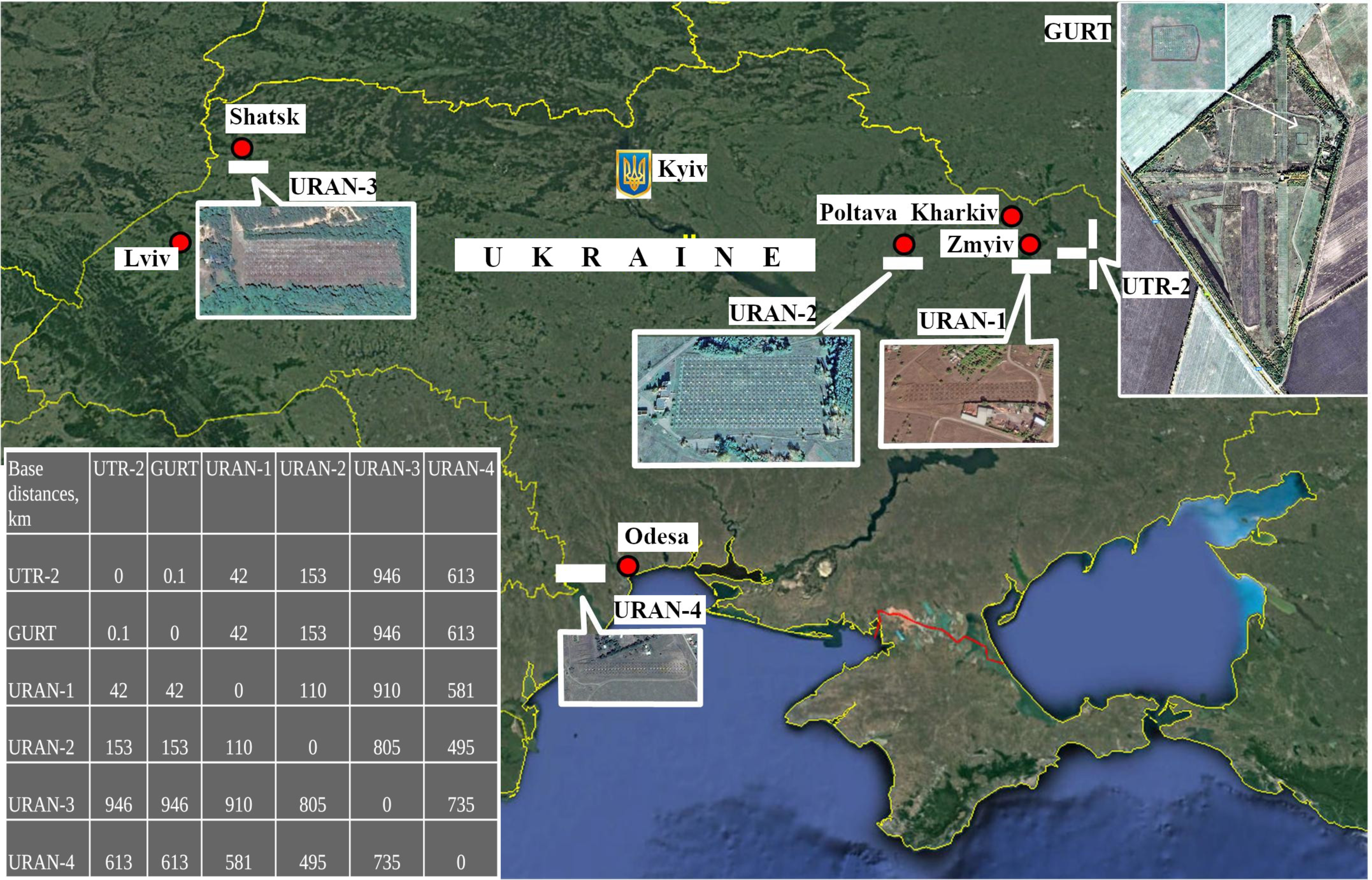}  
\caption{Location of decameter radio telescopes in Ukraine.}
\label{rt4}
\end{figure}

Today, low-frequency radio astronomy is making rapid progress all over the world. New decameter-meter-wave radio telescopes, such as LOFAR, NenuFAR, LWA, MWA, SKA-low, are being built on all continents. However, in terms of their main parameters (frequency band, effective area, sensitivity, resolution, space-frequency response, noise immunity, dynamic range of analog and digital systems), the Ukrainian instruments remain unsurpassed and highly demanded by the international community. They made it possible to achieve a large number of cutting-edge astrophysical results, as well as implement hardware and methodological developments in the course of domestic and international research projects \cite{6}. Figs. \ref{rt1} -- \ref{rt4} show the Ukrainian decameter-wave radio telescopes and their location in Ukraine. The key parameters of the telescopes are shown in Table~\ref{tab1} \cite{2}.

\begin{table*}[t]
\small
\caption{The main parameters of the Ukrainian radio telescopes of the decameter wavelength range.}\label{tab1}
 \begin{tabular}{|c|c|c|c|c|c|c|}
  \hline
           & Frequency & Size, m; & Number of & Beam & Distance & VLBI angular \\
	 Raditelescope; & of range, & maximum & elements & size at & from UTR-2 & resolution \\
	  coordinates & MHz  & effective & $(l\times m=N)$; & 25 MHz  & (LOFAR), km & at 25 MHz \\
		 &   & area, m$^2$ & polarization &   &   & (UTR-2 -- URAN) \\
  \hline
  UTR-2 (RI NASU) & 8--33 & 1800$\times$900; & 2040; & 0.4$^\circ\times$0.4$^\circ$ & 0 & -- \\
 49$^\circ$39$^\prime$N 36$^\circ$56$^\prime$E &  & 140000 & 1 linear &  & ($\approx$2000) &  \\
UTR-2 (North- & 8--33 & 1800$\times$53; & 240$\times$6=1440; & 0.3$^\circ\times$12$^\circ$ & 0 & -- \\
 South arm) &  & 105000 & 1 linear &  & ($\approx$2000) &  \\
UTR-2 (West arm) & 8--33 & 900$\times$45; & 6$\times$100=600; & 0.6$^\circ\times$12$^\circ$ & 0 & -- \\
  &  & 105000 & 1 linear &  & ($\approx$2000) &  \\
	URAN-1 (RI NASU) & 8--33 & 200$\times$29; & 4$\times$24=96; & 5$^\circ\times$30$^\circ$ & 42 & 59$^{\prime\prime}$ \\
 49$^\circ$40$^\prime$N 36$^\circ$21$^\prime$E &  & 5500 & 2 linear &  & ($\approx$1950) &  \\
  URAN-2 (PGO NASU) & 8--33 & 238$\times$116; & 16$\times$32=512; & 3.5$^\circ\times$7.5$^\circ$ & 150 & 16$^{\prime\prime}$ \\
 49$^\circ$38$^\prime$N 34$^\circ$50$^\prime$E &  & 28000 & 2 linear &  & ($\approx$1850) &  \\
URAN-3 (PMI NASU) & 8--33 & 238$\times$58; & 8$\times$32=256; & 3.5$^\circ\times$15$^\circ$ & 946 & 2.6$^{\prime\prime}$ \\
 51$^\circ$29$^\prime$N 23$^\circ$50$^\prime$E &  & 14000 & 2 linear &  & ($\approx$1100) &  \\
URAN-4 (RI NASU) & 8--33 & 238$\times$58; & 8$\times$32=256; & 3.5$^\circ\times$30$^\circ$ & 613 & 4$^{\prime\prime}$ \\
 46$^\circ$24$^\prime$N 30$^\circ$16$^\prime$E &  & 7000 & 2 linear &  & ($\approx$1500) &  \\
GURT, 1 subarray & 8--80 & 18$\times$18; & 5$\times$5=25; & 30$^\circ\times$30$^\circ$ & $\approx$1 & -- \\
 (RI NASU) &  & 650 & 2 linear &  & ($\approx$2000) & \\
49$^\circ$39$^\prime$N 36$^\circ$56$^\prime$E &  & (at 10 MHz) &  &  &  & \\
 \hline
 \end{tabular}
\end{table*}

To properly formulate the requirements for the experimental instrumentation intended to search for the 21~cm hydrogen redshifted line, including the information on its parameters given in the previous chapter, it is required to carefully consider the specifics of decameter wave radio astronomy, in terms of both astrophysical and hardware-methodological parameters. 

One of the key parameters of each radio telescope is its sensitivity. It is determined by the classical ratio \cite{7}
\begin{equation}
 \Delta S_{min}= \frac{2k_B T_{noise}}{A_{ef}\sqrt{\Delta f\Delta t} }\quad \mbox{Jy}\,, \label{dSmin1}
\end{equation}
where $\Delta S_{min}$ is the minimum flux density of electromagnetic radiation detectable from a radio source, $k_B$ is the Boltzmann constant, $T_{noise}$ is the noise temperature of the system, $A_{ef}$ is the effective area of the radio telescope, $\Delta f$ is the frequency bandwidth of the registration (frequency resolution), and $\Delta t$ is the integration or accumulation time (time resolution).

The fundamental difference at low frequencies comparing to other spectral domains is that the noise temperature of the system $T_{noise}$  is not the noise temperature of the receiver $T_{N}$, unlike the higher frequencies (it is about tens of K). At decameter frequencies, it is determined by the brightness temperature of the galactic background $T_{noise} \approx T_{B}(\nu)$, which reaches tens and hundreds of thousands of K. However, despite this huge value, the UTR-2 radio telescope has record-breaking sensitivity due to its gigantic effective area, $\Delta S_{min} \sim 1\,\mbox{Jy} - 1\,\mbox{mJy}$ (depending on the bandwidth and integration time, which in turn are determined by the parameters of the cosmic signals under study).

The above formula and its low-frequency features require a clarification. The approximation $T_{noise} \approx T_{B}$ is only valid when the antenna temperature (a radio signal power from the galactic background) at the output of the antenna or antenna element is much higher than the noise temperature of the receiver (antenna amplifier) connected to the antenna: $T_{a} \gg T_{N}$. However, there are losses in every antenna $\eta_{A}$ (frequency-dependent signal attenuation), including those which use modern broadband active antenna elements \cite{8}. Thus, $T_{a}=T_{B} \eta_{A}$, and the adjusted sensitivity formula is as follows:
\begin{equation}
 \Delta S_{min}= \frac{2k_B T_{noise}}{A_{ef}\sqrt{\Delta f\Delta t} }\left(1+\frac{T_N}{T_a}\right)\,\, \mbox{[Jy]}\,, \label{dSmin2}
\end{equation} 
The sensitivity reduction factor $m=T_a/(T_a+T_N)$ can be significantly less than one. For example, when the antenna background temperature is equal to the receiver noise temperature (sometimes radio telescope designers allow this), the sensitivity is halved, which is equivalent to a halving of the effective area (usually its maximization is the main requirement by radio astronomers). This makes it necessary to introduce a new quality parameter for a low-frequency radio telescope, the System Equivalent Effective Area.
\begin{equation}
 SEEA = A_{ef} T_{a}/(T_{a}+T_{N}) = m A_{ef}\quad \mbox{[m$^2$]}\,. \nonumber
\end{equation}
For low-frequency radio astronomy telescopes, this is a more suitable measure of the instrumental quality compared to the traditional quality parameter at high frequencies is the system equivalent flux density:
\begin{equation}
 SEFD = 2k_B T_{noise}/A_{eff}\quad \mbox{[Jy]}\,. \nonumber
\end{equation}
At low frequencies, $T_{noise}$ is determined primarily by $T_{B}$ and cannot be reduced in any way to improve the quality of the radio telescope compared to higher frequencies (millimeter to decimeter bands),  where $T_{noise}$ can be reduced by using modern semiconductors and cryogenic technology.

Formulas (\ref{dSmin1}) and (\ref{dSmin2}) are used for monitoring of point radio sources with angular sizes smaller than the width of the radio telescope's beam pattern (for UTR-2, the angular resolution is $\sim$30 arcmin). In the case of extended and/or isotropic radio sources (such as the global signal in the 21~cm line of hydrogen from the Dark Ages), it is advisable to use the following expression for the minimum detectable brightness temperature
\begin{equation}
\Delta T_{min}=\frac{T_{B}}{m\sqrt{\Delta f \Delta t}}\,. \label{dTmin} 
\end{equation} 
As can be seen in this case, the effective area, i.e., the size of the radio telescope, is not involved, thus, a small antenna and even a single antenna element with a quasi-isotropic radiation pattern can be used. This conclusion is fundamental in the study of extragalactic nearly isotropic cosmological effects of various types. The fine spatial structure of such sources is studied at the next stages of research after the detection of the effect itself. Maximizing the coefficient $m$ and its approach to 1 remains extremely important in each case. This has been realized for all Ukrainian radio telescopes, as we managed to achieve a very high $T_{a}$ and $T_{N}$ ratio of 6--10 dB, which results in $m=$ 0.8--0.9. Moreover, this is ensured in a wide frequency band of radio telescopes, as shown below.

\begin{figure}[htb]
\includegraphics[width=0.48\textwidth]{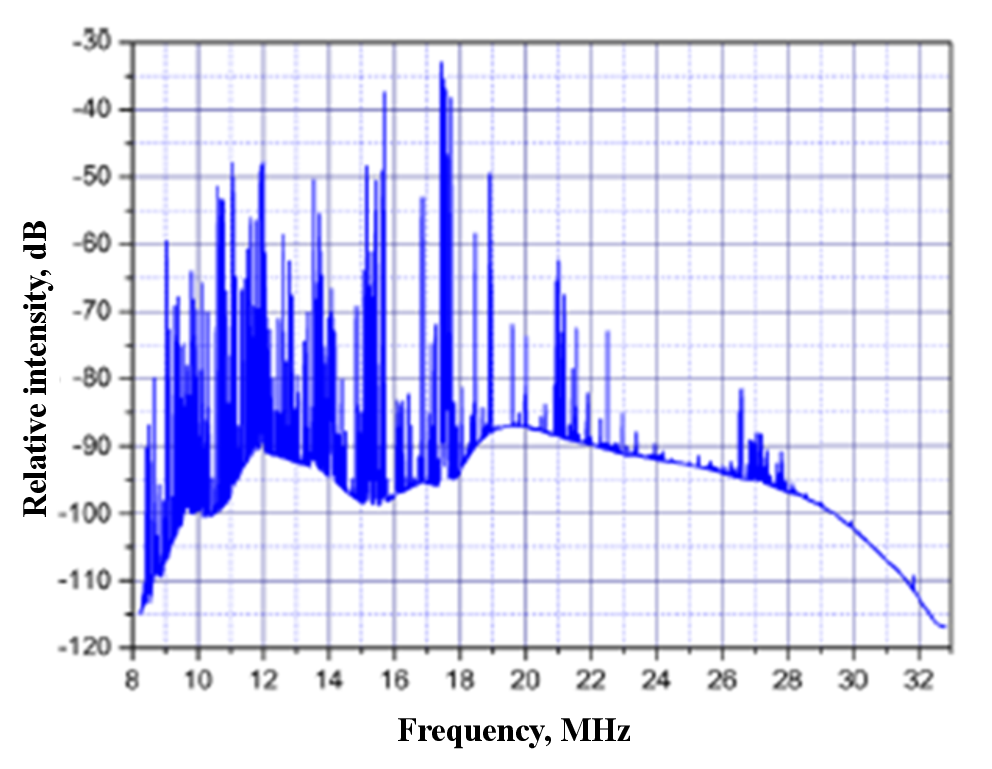} 
\caption{Response of the UTR-2 radio telescope to the radio radiation of the galactic background during the daytime.}
\label{fig5}
\end{figure}

\begin{figure}[htb]
\includegraphics[width=0.48\textwidth]{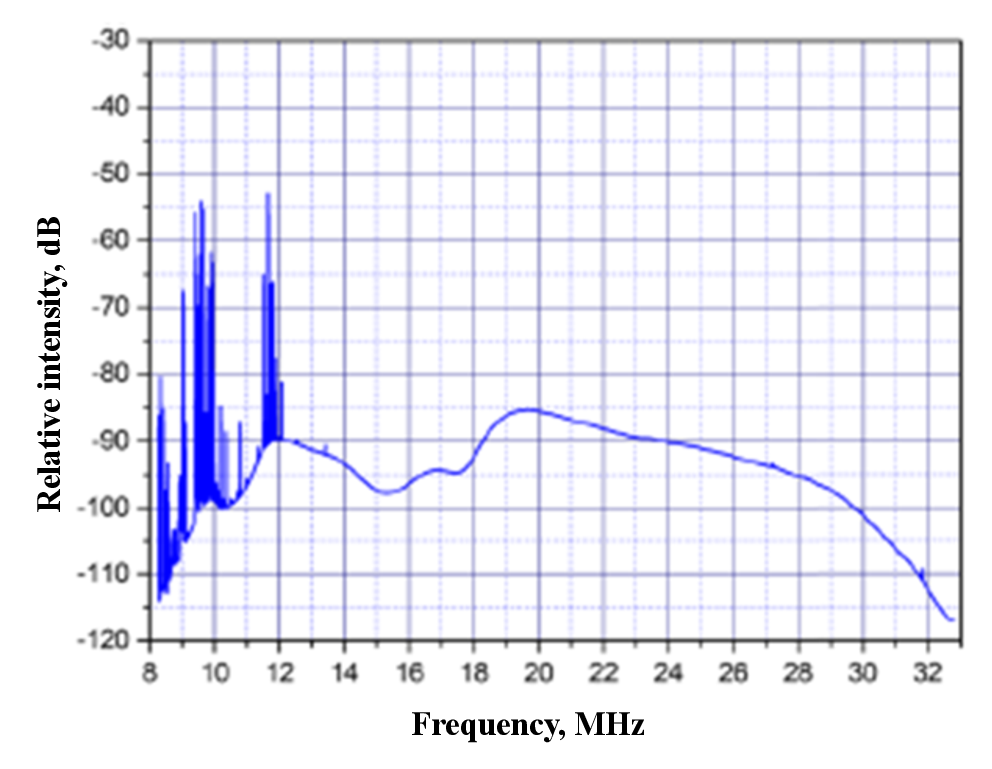} 
\caption{Response of the UTR-2 radio telescope at night.}
\label{fig6}
\end{figure}

Figs.~\ref{fig5}--\ref{fig6} show the amplitude-frequency characteristics (bandwidth) of the UTR-2 radio telescope. These are the real responses of the telescope to the galactic background, which occupies the entire electromagnetic radio spectrum from hectometer to centimeter waves. These responses also contain narrow-band radio interference, the intensity and amount of which significantly depends on the time of day (much less at night) and frequency usually almost absent at frequencies above 15--20 MHz at the UTR-2 site). This radio frequency interference has a width of $\Delta \nu_{RFI} \sim  10 $ kHz. The unevenness of the characteristic is caused by the use of a special 3-band antenna amplification system with the bandwidths of 8--12, 12--18 and 18--33 MHz. This unevenness is corrected in a certain way by calibrations and scaling. The URAN-1 and URAN-4 radio telescopes have similar frequency bands, 8--33 MHz (see Table~\ref{tab1}).

\begin{figure}[htb]
\includegraphics[width=0.48\textwidth]{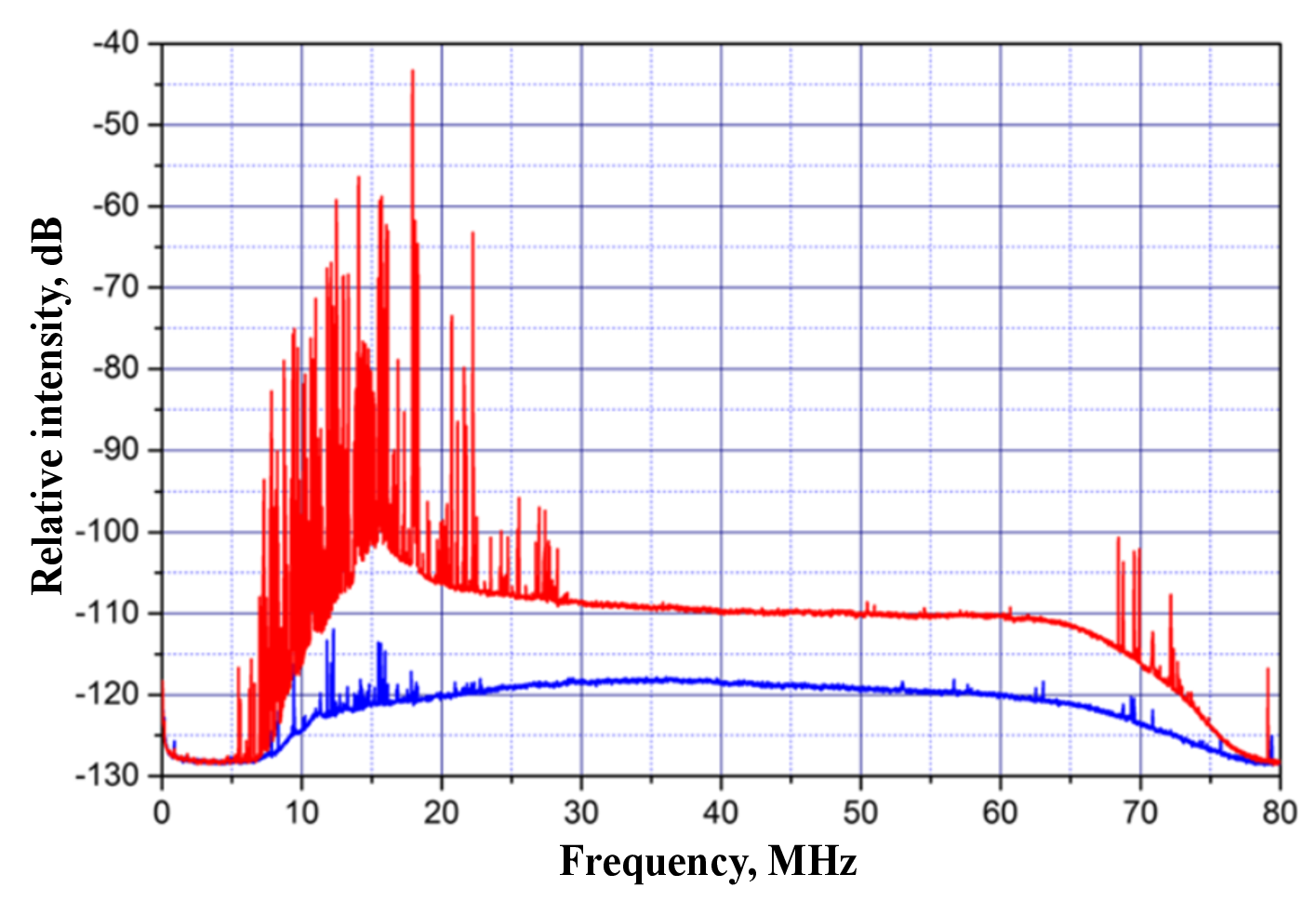} 
\caption{Response of the subarray of the GURT radio telescope to the radio radiation of the galactic background (amplitude-frequency characteristic with the addition of narrow-band radio interference) during daytime. The red line is the antenna temperature, the blue line is the intrinsic noise of the subarray.}
\label{fig7}
\end{figure}

\begin{figure}[htb]
\includegraphics[width=0.48\textwidth]{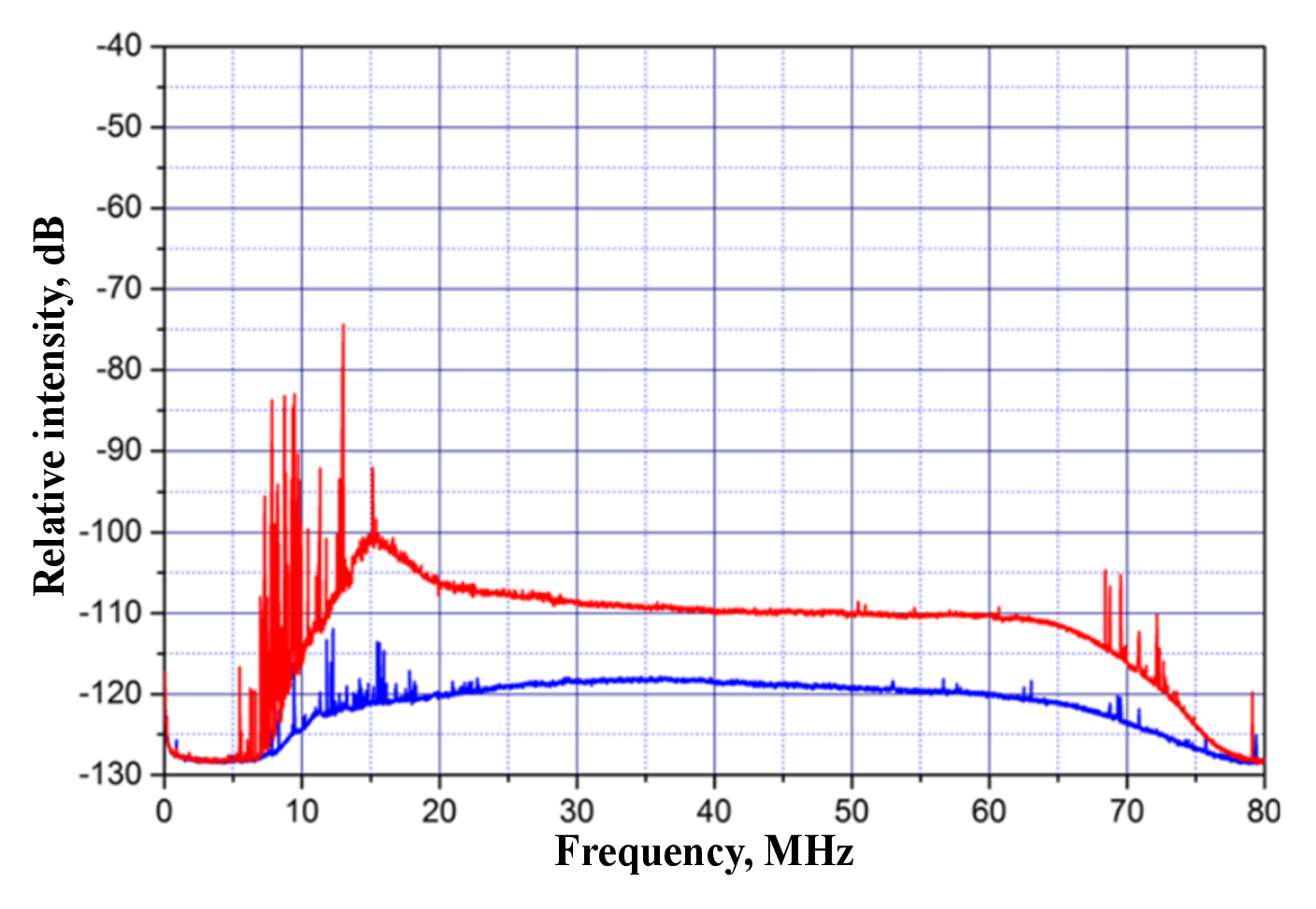} 
\caption{Response of the GURT at night. The red line is the antenna temperature, the blue line is the intrinsic noise of the subarray.}
\label{fig8}
\end{figure}

It is important to emphasize that this frequency band seems to be insufficient to search for very broad cosmological lines of neutral hydrogen (see below). However, a new generation radio telescope, GURT, is being built at the S.\,Y.\,Braude Observatory next to the UTR-2 radio telescope. Despite its smaller size (as shown above in this case, size is not important), this instrument has a number of advantages, among which the main one is a much wider frequency band of 8--80 MHz. The GURT telescope is assembled from 25-element bipolarized subarrays with analog phasing in the subarray and digital phasing between them. Figs.~\ref{fig7} -- \ref{fig8} show the amplitude-frequency characteristics of the GURT in the daytime and at night. A much greater uniformity of the transmission coefficient is also seen, which is useful, too. 

Besides, the lower lines in both figures show the intrinsic noise level of the subarray, which is $\sim$10~dB lower than the antenna background temperature, giving a very high value of $m\approx 0.9$ \cite{9}.

Thus, the expediency of using GURT subarrays and even their individual elements is obvious. They provide the required frequency band, sensitivity, uniformity of the transmission coefficient, stability, noise immunity, cost-effectiveness and efficiency of operation. It should be recalled that the subarrays and antenna elements of the LOFAR (the Netherlands and other European countries) and NenuFAR (France with the participation of Ukrainian radio astronomers) systems have similar frequency bands of 10(20)--80 MHz and can also be used in a coordinated manner to search for redshifted HI lines at decameter-meter waves.

It should be emphasized that the appearance of the Universe at the lowest wavelengths that can be observed from the Earth surface (i.e. the decameter range) is very different from that of, for example, optics or high-frequency radio astronomy. Non-thermal mechanisms of radio emission, non-stationary nonequilibrium radio emission during the propagation of waves and charged particles in a magnetically active plasma, prevail at decameter wavelengths. There is a strong interaction of radio radiation with the plasma (absorption, refraction, scattering), large-scale evolution and subtle atomic processes are manifested. Thus, low-frequency radio astronomy allows obtaining unique information that is not available to the other methods of astrophysical research.

\begin{figure*}[htb]
\includegraphics[width=0.98\textwidth]{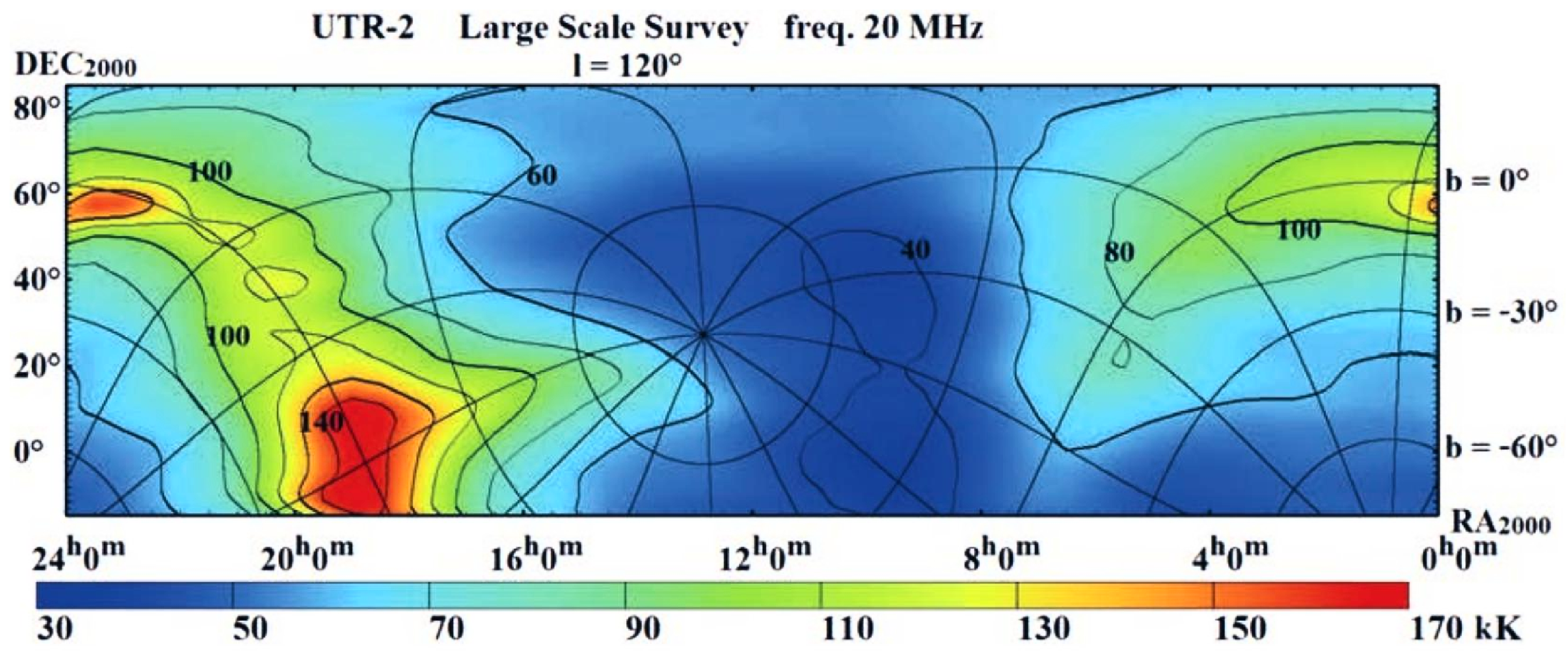} 
\caption{Map of radio emission of the galactic background at a frequency of 20 MHz \cite{26}.}
\label{fig9}
\end{figure*}

\begin{figure}[htb]
\includegraphics[width=0.48\textwidth]{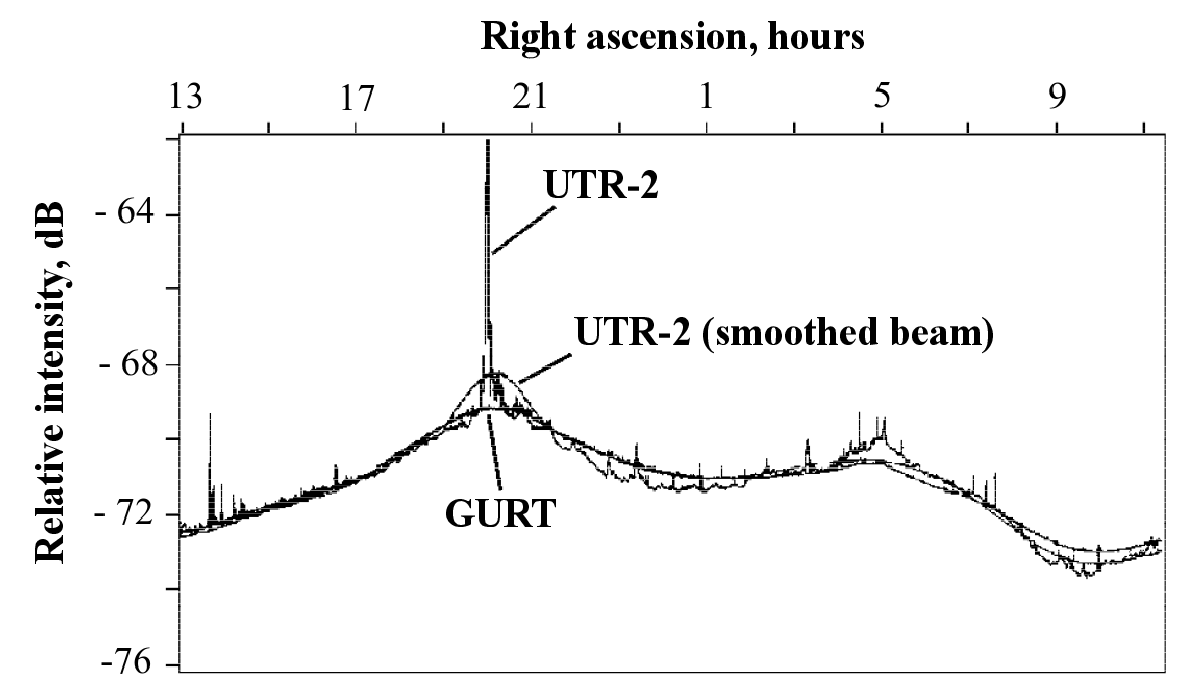} 
\caption{Daily scan of the radio emission of the Galaxy with stationary directional diagrams of the East-West UTR-2 antennas and the GURT subarray at a frequency of 25 MHz. The intense peak corresponds to the radio source Cyg~A ($\delta=40^\circ$ 30').}
\label{fig10}
\end{figure}

An important example of a low-frequency astrophysical phenomenon is the non-thermal synchrotron radio emission of the Galaxy, which has already been mentioned above. It is generated by the motion of relativistic electrons (one of the components of cosmic rays) in the magnetic field of the Galaxy and is very intense at decameter wavelengths: it reaches an average of $\sim 5\times 10^5$~К at a frequency of 10 MHz and $\sim 2\times 10^4$~K near~30 MHz. The emission spectrum has a power-law form of the $T_{B} (\nu) \propto  \nu^{-2.6}$ type.  Fig.~\ref{fig9} shows a large-scale map of this galactic background, constructed at a frequency of 20 MHz (20~kHz registration bandwidth) by the combined use of the UTR-2 section and the URAN-2 antenna (angular resolution of about $10^\circ$). The principal feature is a clearly visible change in the background luminosity temperature depending on the coordinates (galactic or equatorial). Thus, near the Galactic plane relative to the poles, the change in the $T_{B}$ temperature reaches $\sim$6 dB. Fig.~\ref{fig10} shows the diurnal change in the intensity of the signal received when the GURT and UTR-2 antennas, with the beams stationary in the meridional direction, are oriented towards declination $\delta \approx 40^\circ$, which corresponds to the radio source Cyg~A. The intersection of the map in Fig.~\ref{fig9} for the corresponding declination in the direct ascension interval of $\alpha=0^{h}-24^{h}$ is ensured by the diurnal rotation of the Earth.

\begin{figure*}[htb]
\includegraphics[width=0.98\textwidth]{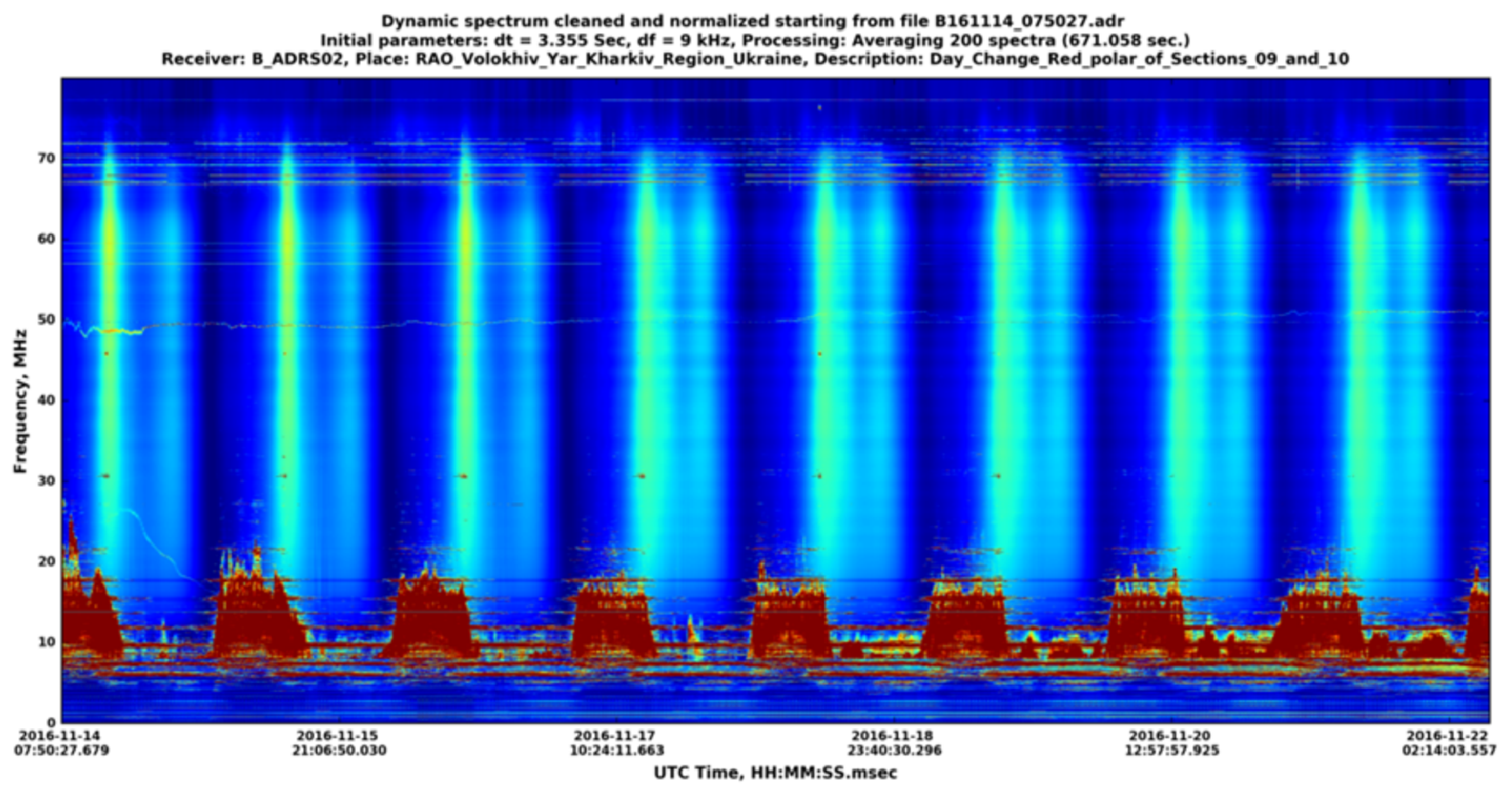} 
\caption{Eight-time passage of the Galaxy through the fixed directional diagram of the GURT subarray, which is oriented at the inclination of $\delta=50^\circ$ (the zenith position of the diagram).}
\label{fig11}
\end{figure*}

Fig.~\ref{fig11} shows a similar experiment when the beam of the GURT subarray is oriented at the zenith $\delta = 50^\circ$.  The result of 8 days of scanning is presented in three-dimensional form: time – frequency – intensity (brightness shown: light color -- maximum intensity, dark color -- minimum). It can be seen that over the entire frequency band of 8--80~MHz, the intensity change is not less than 3 dB and is repeated from day to day with high accuracy.

It is important to note that in the case of such diurnal observations, but with one antenna element only (having a very wide diagrammatic directivity of $\theta_{A} \gtrsim 100^\circ$), a change in the signal intensity from the galactic background of $\sim$3 dB (twofold) was also recorded. This was confirmed by measurements on individual active antenna elements of the GURT and on the dipoles of the UTR-2. The same was shown in the early 1970s during precision experiments on half-wave antenna dipoles of decameter waves \cite{10}. The corresponding diurnal variation of intensity is useful when implementing the methodology of searching for faint lines, both using arrays of antennas and for individual elements. In the latter case, low angular resolution, as shown above, is acceptable, since the radio source under study is isotropic in space.

The galactic background described above sets fundamental restrictions for the study of extragalactic cosmological effects at decameter and meter wavelengths (this is the so-called ``foreground''). In terms of its luminosity temperature, it is tens and hundreds of thousands times higher than the temperature of CMB $T_{\rm CMB}=2.73$~K, which is the ``background'' in the formation of the HI lines in the absorption for $z \sim  80$.  Thus, the relative intensity of the lines to be searched for should be estimated in comparison with the Galactic background, which drastically reduces the observability of lines in real-world experiments.

\section{Radio spectroscopy at decameter wavelengths} 
 
One of the key achievements of decameter radio astronomy is the detection of an exotic astrophysical phenomenon, i.\,e. spectral lines of extremely excited interstellar carbon atoms for principal quantum numbers $n>600\, (\nu < 30\, \mbox{MHz})$. This discovery was made about 45 years ago 
using the UTR-2 radio telescope \cite{11}. The corresponding recombination lines were first observed in the absorption against the background of strong non-thermal radio radiation. This absorption phenomenon provides additional advantages in the reliable identification of a faint effect and must be considered when searching for new absorption lines. The detected lines have become a new effective means of diagnosing cold partially ionized interstellar plasma, and the Ukrainian instruments remain the most convenient for the development of this promising methodology \cite{12,13}. It is no coincidence that the largest number of objects in the Galaxy in the decameter wavelength range has been studied using the UTR-2, which is important for future observations of the other lines in the Galaxy and from extragalactic sources \cite{14,15}. Besides, it is important to know the main parameters of low-frequency recombination radio lines and their influence on the search for cosmological lines of neutral hydrogen.

Recombination line frequencies are determined using the Rydberg formula
\begin{equation}
\nu_{RL}=c Z^{2} R\left(1-\frac{m}{M}\right)\left[\frac{1}{n^{2}} -\frac{1}{(n+\Delta n)^{2}}\right]\,, \label{nuRL} 
\end{equation}
where $R$ is the Rydberg constant, $m,\,M$ are the masses of the electron and the atom, respectively, $c$ is the speed of light, $Z$ is the nucleus charge, $n$ and $\Delta n$ are the principal quantum number and its change, respectively. It is easy to show that on decameters, the distance between adjacent lines is $(n>>\Delta n\sim 1)$: 
\begin{equation}
\Delta \nu_{RL} \approx \frac{3 \nu_{RL}}{n}\,. \label{dnuRL}
\end{equation}
This creates a series of lines corresponding to different $n$ and $\Delta n$. Independent series appear for different atoms (isotopic shift) according to factor $(1-m/M)$. Let us take an example of the line of a single-ionized carbon ($Z = 1$), which appears as a result of the electron transition between adjacent levels with principal quantum numbers 640 and 639 (lines appearing from $\Delta n = 1$ transitions are called $\alpha$-lines). The transition frequency according to (4) is $\nu_{RL} (640) \approx 25$ MHz, and the distance to the adjacent line ($n=641\,\rightarrow\,n=640$) according to (5) is $\Delta \nu_{RL} \sim  125$ kHz.

The integral relative intensity of the lines is defined as \cite{12}
\begin{equation}
I_{RL} = \int \Delta T_{RL}/T_{B} d\nu \approx 2 \cdot 10^{6} \frac{N_e^{2}\,l\,b_n\beta_n}{T_{e}^{5/2}}, \mathrm{c}^{-1}\,, \label{IRL} \nonumber
\end{equation} 
where $\Delta T_{RL} = - T_{B} \tau_{RL}$ is the absolute intensity at the optical depth of $\tau_{RL}$, $N_{e}, T_{e}$ are the electron density and temperature of the interstellar cloud, $l$ is its length along the visual path, and $b_{n},\,\, \beta_{n}$ are the disequilibrium factors determined by the population of atomic levels. 
  
The corresponding lines’ width is determined by the Doppler $\Delta \nu_{D}$, Stark’s $\Delta \nu_{P}$  and radiative $\Delta \nu_{R}$ mechanisms and is equal to \cite{12}
\begin{equation}
\Delta\nu_{DPR} = \sqrt{\Delta \nu^{2}_{D} + (\Delta \nu_{P} + \Delta \nu_{R})^{2}}\,. \label{dnuDPR} \nonumber
\end{equation} 

Currently it is well established that decameter carbon recombination lines appear in partially ionized cold diffuse interstellar hydrogen clouds when the latter is virtually neutral and carbon is fully ionized \cite{15}.  This is due to different ionization potentials: $E_{H} = 13.6$ eV,  and $E_{С} = 11.2$ eV (the main source of ionization is ultraviolet quanta with a wavelength of 913 \AA $< \lambda < 1100$ \AA). The physical conditions in such clouds are as follows: electronic temperature $T_{e} \sim  50-100$ К, electronic concentration $N_{e} \sim  1-0.001\,\, \mbox{cm}^{-3}$, size $l \sim  1-100\,\, \mbox{pc}$, turbulent motion $\Delta V_{T} \sim  10 $ km/s (1 kHz), and maximum radial velocity $V_{r} \sim  300$ km/s (30 kHz). In this case, the key parameters of the decameter lines, which are confirmed by experiments and calculations, are as follows: line spacing $\Delta_{RL} \approx 120$ kHz, the number of $\alpha$-lines within 8--33 MHz interval, $N_{RL} \approx 300$, 
line width $\Delta \nu_{DPR} \approx 1-10 $ kHz, relative intensity $\Delta T_{RL}/T_{B} \sim  10^{-3} -10^{-5}$, and integral relative intensity $I_{RL} \sim  10-10^{-2}$ c$^{-1}$.

\begin{figure}[htb]
\includegraphics[width=0.48\textwidth]{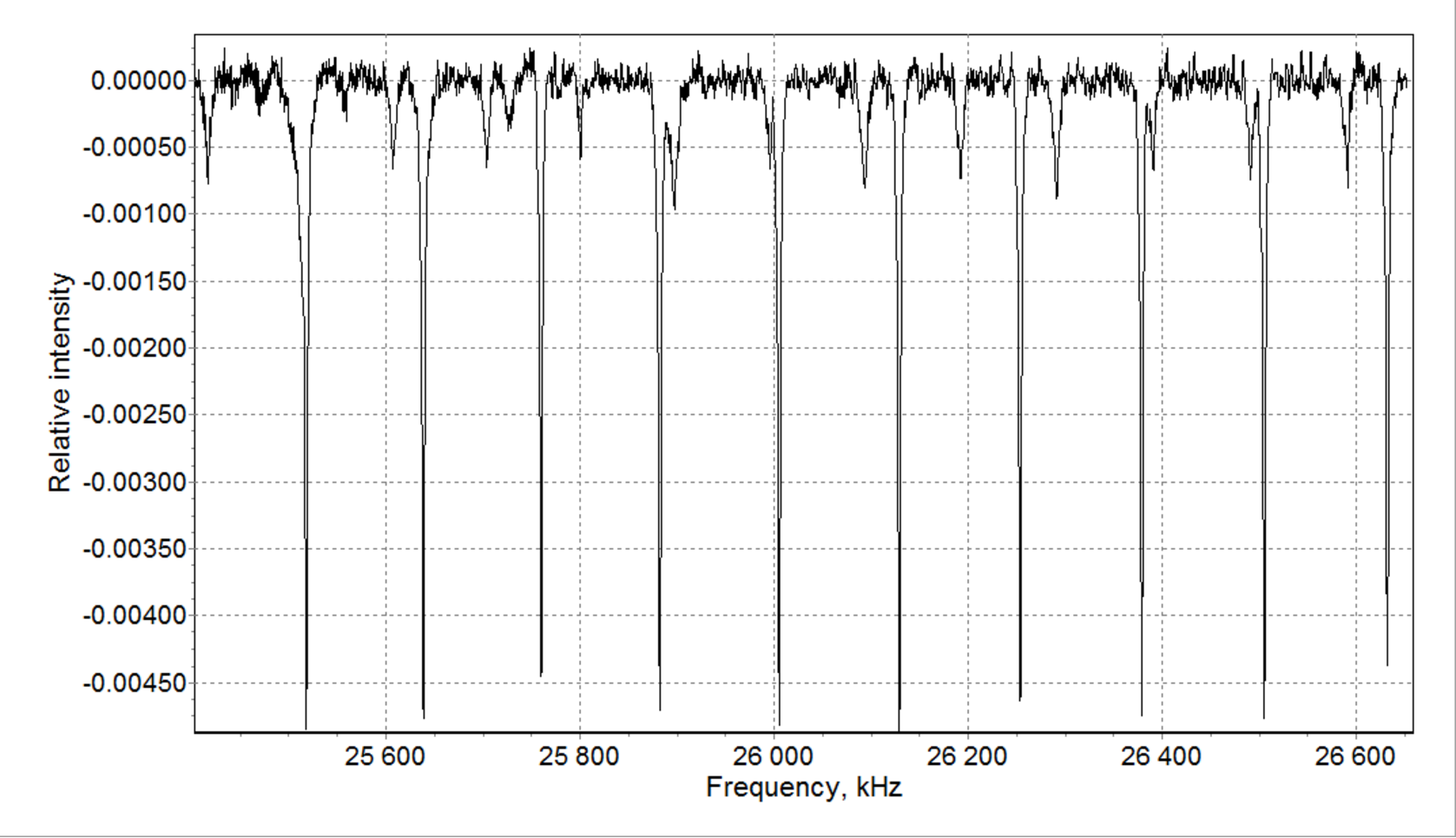} 
\caption{High sensitive simultaneous observations of ten recombination radio lines of carbon C 627$\alpha$ -- C 636$\alpha$ near the frequency of 26 MHz.}
\label{fig12}
\end{figure}

\begin{figure}[htb]
\includegraphics[width=0.48\textwidth]{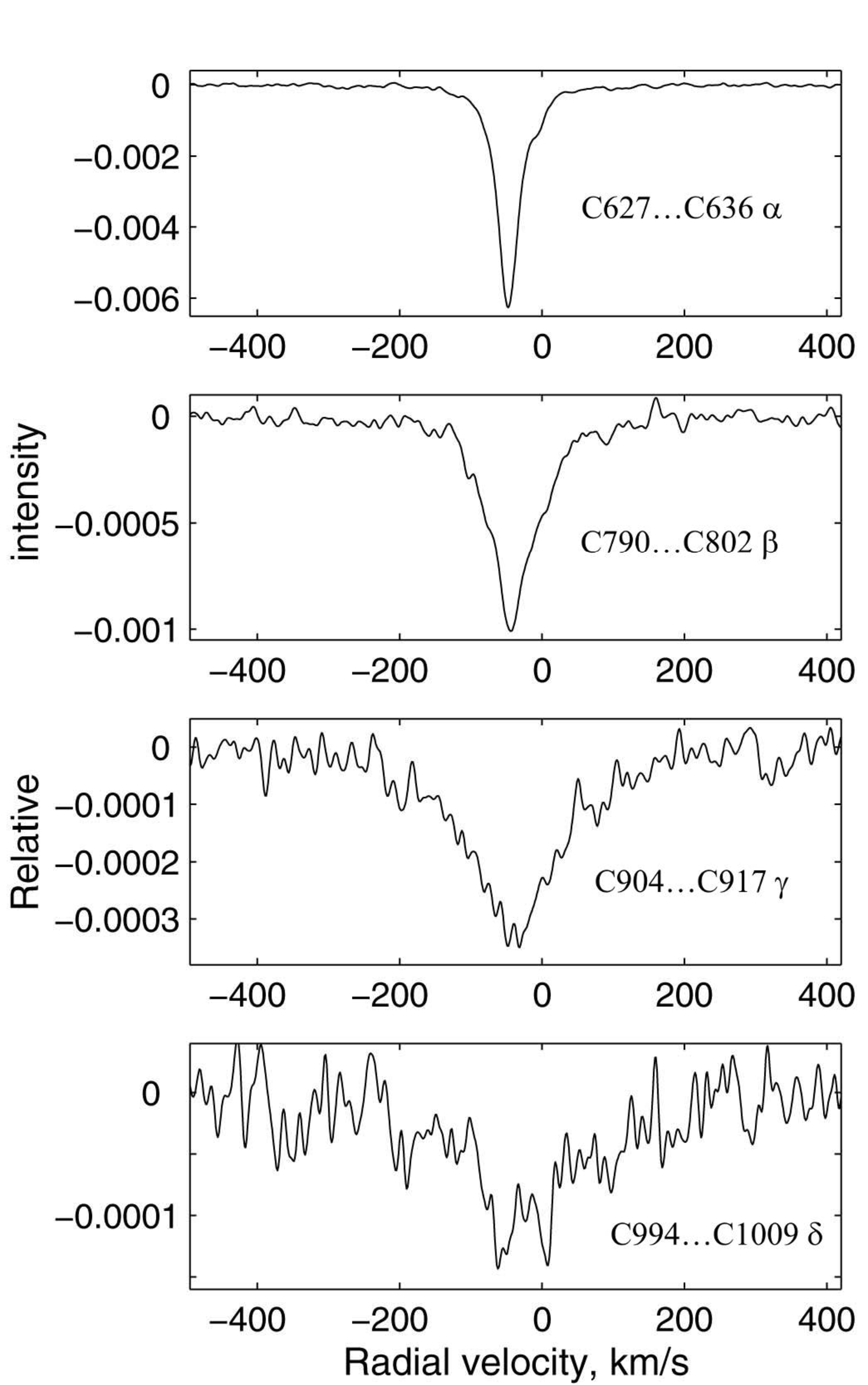} 
\caption{Detection of $\alpha$, $\beta$, $\gamma$, $\delta$ - recombination radio lines and discovery of record highly excited states of interstellar atoms at $n\sim1000$ levels.}
\label{fig13}
\end{figure}

To confirm the above, Figs. \ref{fig12}, \ref{fig13} show an example of detecting recombination lines at a frequency of $\sim  26$ MHz in the direction of the Cas~A radio source. The parameters of the lines are quite typical for other Galaxy objects.

Thus, it is clear that there is a large number of recombination lines in the Galaxy over the entire spectrum of decameter waves \cite{15,16}. They are concentrated mainly near the Galactic plane, but also observed at rather large galactic latitudes $|b|>10^\circ$. Despite relatively low intensities of these lines, they are still significantly higher than the absolute and relative intensities of the cosmological HI lines that are proposed to be searched. Nevertheless, the extensive knowledge of the parameters of the low-frequency recombination lines allows us to expect their effective separation from the radio spectroscopic observational effect associated with the HI absorption line shifted towards decameter wavelengths.

As it follows from Section I and the related article \cite{Novosyadlyj2023a}, the key parameters of the expected line are as follows:
\begin{itemize}
\item the frequency of the 21~cm hydrogen line within its own reference frame $\nu_{H}=1420$ MHz, 
redshift $z\approx  87$, 
\item frequency of the line shifted by the Universe expansion for an observer on Earth $\nu_{L}\sim 16$ MHz, 
intensity (average) $\Delta T_{L}\approx  0.04$ К\footnote{In section І, in the related article \cite{Novosyadlyj2023a} and in \cite{Novosyadlyj2020a,Novosyadlyj2023}, the differential brightness temperature in 21 cm lie of the neutral hydrogen is labeled as $\delta T_{br}$, hereinafter $\Delta T_{L}$}, 
\item the width at the level of 0.5 of the maximum $\Delta \nu_{L}\approx  25 $ MHz,  
\item the width at the level of 0.1 of the maximum $\Delta \nu'_{L}\approx  50$ MHz, 
\item the polarity is negative, without polarization, 
\item the angular size of the source is $\Theta_{S} = 4\pi$ (isotropic), 
\item the background temperature is $T_{CMB}(0) \approx 2.73$~K, and 
\item the foreground temperature near 20 MHz $T_{B}$ is $4\times 10^4$~K.
\end{itemize}.

Thus, in view of the above line parameters, radio spectroscopic requirements, and the experience of previous research at decameter wavelengths, we are going to formulate the key parameters of the experiments as follows:
\begin{itemize}
\item The analysis bandwidth         $\Delta F=(2-10) \Delta \nu_{L} \approx 50-250$~MHz
\item Frequency resolution  $\Delta f = \Delta \nu_{obs}/(1-10) \approx 25-2.5$ MHz
\item The number of frequency channels   $M=\Delta F/\Delta f_{min} = 20-100$
\item The sampling frequency        $F\geq 2\Delta F= 100-500$ MHz
\item Sampling quantization    $q=16$ bit
\item Spectral resolution adjusted to the width of radio interference or recombination lines $\Delta f_{RFI} = 1-10$ kHz
\item The number of channels          $M_{RFI} = 5000-25000$
\item Angular resolution $\Theta_{A}= 30^\circ - 180^\circ$
\item Polarization                   $N_{p} = 2$
\item Relative sensitivity $(T_{B}\gg T_{CMB})$    $\Delta T_{L}/T_{B} \approx 10^{-6}$
\item The signal to noise ratio     $S/N \rightarrow 10$
\item The fluctuation level at spectra\\  $\sigma\leq (\Delta T_{L}/T_{B})/(S/N) = 10^{-7}$
\item The time-based resolution  is   not required (the effect is stationary)
\item The spectrum analyzer type a) digital autocorrelating, b) digital with direct fast Fourier transform.
\end{itemize}

Let us 
draw attention to some important requirements for an experiment in the field of low-frequency radio astrospectroscopy.

The analysis bandwidth should be several times larger than the line width. This ensures applicability of relative spectroscopic measurements, which are much more accurate than absolute ones. The intensity in a line is compared to the neighboring spectrum level where lines are known to be absent, which is why we choose a GURT antenna with a bandwidth of 8--80 MHz. Estimates show that it is almost impossible to achieve a much wider bandwidth while maintaining the maximum sensitivity $m \sim  0.9$ and noise immunity. Antenna elements on the far side of the Moon will improve the situation in the future.

The resolution must be several times better than the line width in order to determine the line shape (the number of channels is several dozens).

In the presence of narrow-band radio interference inherent at decameter wavelengths, the resolution should be quite high $(\leq 10\,\, \mbox{kHz})$, and the number of spectral channels should reach many thousands. The contradiction of the latter two requirements is resolved by using a higher resolution at the observing stage, and during the secondary processing of the obtained spectra with interferences and recombination lines, the latter are eliminated by special digital filtering, and then the spectra themselves are ``smoothed'' in terms of frequency and according to the expected width of the cosmological lines.

Under these conditions, the key parameter of faint-line radio spectroscopy is the required $\Delta t$ accumulation time at a given sensitivity and resolution. Then, according to (3), we get 
\begin{equation}
\sigma = \Delta T_{min}/T_{B} = 1 /\sqrt{\Delta f \Delta t} = 10^{-7}\,. \label{tau}
\end{equation} 
For $\Delta f= 2.5$ MHz, it leads to
\begin{equation}
\Delta t = \frac{1}{\Delta f \sigma^{2}} = 463\quad \mbox{days}\,. \label{dt}
\end{equation}

\begin{figure*}[htb]
\includegraphics[width=0.95\textwidth]{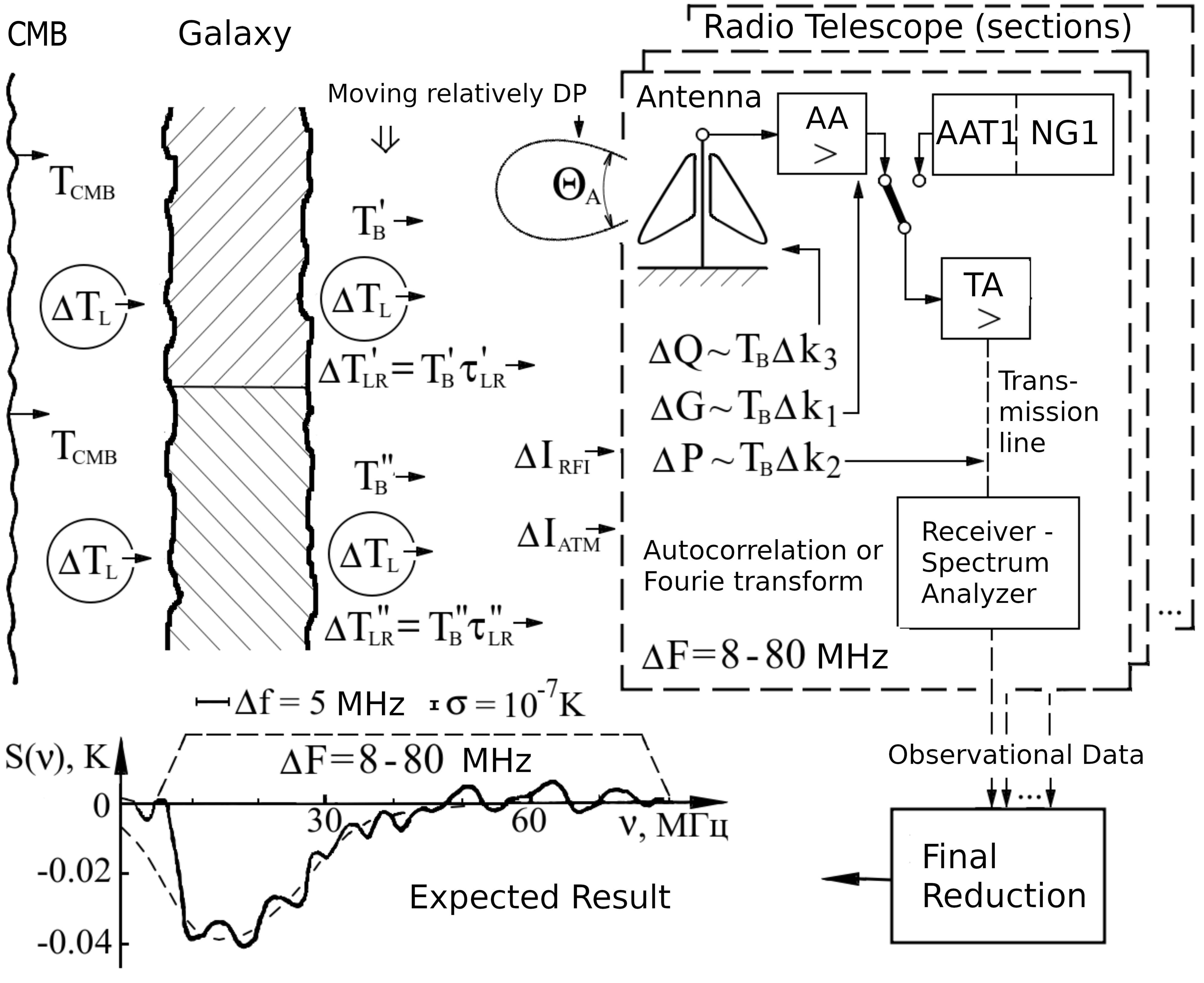} 
\caption{Structure of the experiment.}
\label{fig14}
\end{figure*}
 
This is an extremely long period of continuous observation ($\approx 1.3 $ yr), which is impossible to achieve in practice.

However, $\Delta t$ can be reduced significantly under some of the above considerations. If we neglect the determination of the line shape and focus on the fact of detection itself (narrow-band interferences have already been eliminated), then during secondary processing we can further smooth out the spectra with a frequency resolution of $\Delta f = \Delta \nu_{L} = 25 $ MHz (this is optimal from the theoretical point of view on the effect detection). As can be seen from (\ref{dt}), the integration time is reduced by a factor of 10 and becomes quite realistic: 
\begin{equation}
\Delta t = \frac{1}{25 \cdot 10^{6} \times 10^{-14}}\,\,\mbox{sec} \approx 46\quad \mbox{days}\,. \nonumber
\end{equation}
Unfortunately, there are additional interfering factors that greatly complicate the experiments aiming to search for extremely faint and broad enough extragalactic spectral lines, which will be discussed in the following sections.
 
\section{Interfering radio spectroscopic factors and methods of searching for cosmological hydrogen lines at decameter wavelengths}

Fig.~\ref{fig14} shows the structure of the proposed experiment. 
Extragalactic regions of the visible Universe containing astrophysical objects of research, i.e. cosmic microwave background with a brightness temperature of $T_{CMB}\approx 2.73$~K, against which the hydrogen absorption line with intensity of $\Delta T_{L}$  is generated and the search for which is the main goal of the experiment, are shown on the left side of the figure.  Much closer is the Galaxy with powerful non-thermal radio emission, which has a continuous spectrum with a significant gradient, fills the entire sky, but has an intensity change of several dB $T'_{B}$ and $T''_{B}$, depending on the galactic coordinates. The low-frequency narrow-band recombination radio lines described in the previous sections are formed in the Galaxy, with intensities $\Delta T_{LR}'$ and $\Delta T''_{LR}$, which depend on the background radio sources $T'_{B}$ and $T''_{B}$ and physical conditions along the path of the optical depths $\tau'_{LR}$  and $\tau''_{LR}$. The right section of the figure shows a simplified diagram of the radio telescope, consisting of the antenna system itself, antenna and trunk amplifiers (AA, TA), signal transmission lines, spectrum analyzers, computers for data recording and processing. This leads to the final result shown in the lower left section of the figure.
The antenna directional pattern (DP) with the width of $\Theta_{A}$ is displaced relative to the Galaxy due to the diurnal rotation of the Earth. In the near-Earth space, there is a narrow-band radio interference from $\Delta I_{\rm RFI}$ radio stations (mentioned in the previous section), as well as relatively broadband interferences arising from certain atmospheric (more precisely, ionospheric-magnetospheric) phenomena.

As can be seen from the above, only one of the factors shown in the figure is positive and related to the subject of the study, 
the cosmological line of neutral hydrogen, which is shifted towards the decameter range of radio waves and has a negative intensity $\Delta T_{L}$. All other parameters and phenomena shown in the figure and received (or generated) by the radio telescope are interfering and significantly complicate the solution of the problem. Let us analyze them in more detail, since overcoming them is the main goal of the observational methodology proposed in this paper.

\subsection{Free-free absorption}

Free-free absorption of $T_{B}$ and $\Delta T_{L}$ radio radiation is generated in the plasma environment of the Galaxy. As can be seen from Fig.~\ref{fig14}, on the way of extragalactic radio radiation propagation towards the observer, there is a thick layer of our Galaxy with an approximate size of 20 kpc $\times$ 1 kpc, the main component of which is the fully or partially ionized interstellar medium. One of the phenomena in this environment is the free-free absorption of radio radiation, which increases significantly as the frequency is decreasing. Such an absorption, for example, for $\Delta T_{L}$,  is described by the following expression
\begin{equation}
\Delta T'_{L} = \Delta T_{L}\exp (-\tau_{c})\,, 
\label{DTL}
\end{equation}
where $\Delta T_{L}$ is the brightness temperature of the radio radiation before it penetrates the Galaxy, 
$\Delta T'_{L}$ is the brightness temperature ``at the output'', and $\tau_{c}$ is the optical depth of free-free absorption in the plasma. The latter is defined as \cite{7}
\begin{equation}
\tau_{c}= \frac{0.0314 N_e^{2} l}{\nu^{2} T^{3/2}_{e}}[1.5 \ln{T_{e}} - \ln{(20.2 \nu)}]\,, 
\label{tauc}
\end{equation}
where  $\nu$ is the observation frequency in GHz, $T_{e}$ and $N_{e}$ are the electron temperature and density, respectively, and $l$ is the length of the medium along the line of sight.

For emission nebulae (such as the Rosette, North America, and other $HII$ regions) at 20~MHz, the optical depth is very large $(\tau_{c} \gg 1)$ at electron temperatures and densities $T_{e} \sim  10^{4}$ К, $N_{e}= 10 - 10^{3}\,\, \mbox{cm}^{-3}$. The same happens in partially ionized cold plasma at the periphery of high-density molecular clouds $T_{e} \sim 20$~K, $N_{e} \sim 1\,\, \mbox{cm} ^{-3}$. Consequently, there is almost complete absorption of low-frequency radio radiation, as follows from (\ref{DTL}) and (\ref{tauc}). However, these interstellar formations account for only about 10\% of the total Galaxy volume. The lion's share of the interstellar medium is made up of diffuse neutral hydrogen clouds $(T_{e} \sim 100$~K, $N_{e} \sim 0.001\,\, \mbox{cm}^{-3})$, and the intercloud and inter-arm medium with $T_{e} \sim 10^{4}$ K and $N_{e} \sim 0.01\,\, \mbox{cm}^{-3}$. Even at a distance along the line of sight of about~10 kpc, according to (\ref{tauc}), we have $\tau_{c} = 0.1 - 0.001$. This gives a very low absorption in the Galaxy when observing almost isotropic radio emission $\Delta T_{L}$ with a relatively wide radiation pattern of the radio telescope. The above is also confirmed by the detection of many thousands of discrete radio sources at decameter wavelengths using the UTR-2 \cite{17}. Spectra in the frequency range of 10–25 MHz show a decrease in flux density at the lowest frequencies only in some cases. This is true for both the most powerful galactic radio sources Cas~A, Tau~A, and the radio galaxies Cyg~A and Vir~A, which have linear spectra in the logarithmic scale up to a frequency of 10~MHz.

\subsection{Foreground}

There is a high brightness temperature of the Galactic backgrounds $T'_{B}$ and $T''_{B}$ and the significant frequency gradient $T_{B}\propto \nu^{-2.6}$ (see Section ІІ). Since it is fundamentally impossible to reduce these parameters, the only method to combat the drastic decrease in the relative intensity of the effect being sought $\Delta T_{L}/ T_{B} < 10^{-6}$, is to significantly increase the observation time (up to many months) and use some hardware and software implementations (multi-antenna observations, effective zero line correction).

\subsection{Narrow-band radio interference}

Narrow-band radio interference $\Delta I_{RFI}$ from radiation generated by radio stations, special service radio systems, parasitic radio emissions from the domestic radio telescope instrumentation (computers, monitors, generators, power supplies and so on). This is inherent at decameter wavelengths, but many years of research in this domain allowed us to develop methods for monitoring the interference, improving electromagnetic compatibility of equipment, and implementing criteria for identifying useful signals and interference, including complete elimination of the latter. In the case of the search for cosmological hydrogen lines, the key identification criterion is the fundamental difference in the frequency width of the effects for $\Delta T_{L} (\Delta \nu_{L} \sim 25\,\, \mbox{MHz})$ and $\Delta I_{RFI}(\Delta f_{RFI} = 1-10\,\, \mbox{kHz})$. It seems very promising to use machine learning methods to solve the problems of combating such and other types of interference in low-frequency radio astronomy. In any case, high frequency resolution at the observation stage, $\Delta f_{RFI} =1-10\,\, \mbox{kHz}$, appropriate noise removal and frequency averaging up to $\Delta f = 2.5-25\,\, \mbox{MHz}$ are required.

\subsection{Low-frequency radio interference of natural origins}

Low-frequency radio interference $\Delta I_{ATM}$ is associated with phenomena in the Earth lower atmosphere, ionosphere and magnetosphere. A typical example of natural impulsive radio interference in low-frequency radio astronomy is the short-time radio emission that occurs during electrostatic discharges in the atmosphere, such as lightning. Although this effect is quite powerful, at first glance it does not seem to be dangerous, as the time width is very small -- a few milliseconds, and the spectrum is quite broad, continuous, without spectral features. However, this is true for ``close'' lightning. Due to the ionospheric propagation of radio waves from ``distant'' discharges, spectral features with a width of tens to hundreds of MHz can appear, which is dangerous because it is similar to the spectral width of the effect being sought. Recently, a similar interference effect was detected during observations using UTR-2. It is likely to be caused by the superposition of signals from a large number of lightning bolts, which is typical for tropical latitudes (about 100 lightning bolts per second, the interval between them being $\sim 10 $~ms). Reflecting from the ionosphere at certain frequencies $(\lesssim 15\,\, \mbox{MHz})$ the signals arrive at the radio telescope at angles $\Delta \approx 10^\circ-40^\circ$ above the horizon and create additional relatively broadband power within the range of $\sim 8-15$ MHz, which is dangerous because it falls in the range of parameters of the expected HI line. However, it was found that a similar broadband spectral effect were observed at low time-dependent resolution $\Delta t_{p} > 100$ ms. At a high resolution of $\Delta t_{p} = 1-10$ ms, the quasi-continuous effect in terms of time and frequency is divided into separate pulses, which are much easier to filter out against the background of the time-stationary broadband isotropic phenomenon of the cosmological hydrogen line. Furthermore, observations in the zenith direction, especially those with a moderately narrow radiation pattern, will significantly reduce the aforesaid interference effect.

In addition to the above, there are some other atmospheric and magnetospheric frequency-dependent radio physical phenomena (auroral emission, Schumann resonances, etc.). However, all of them are concentrated at extremely low frequencies of $\ll 1$ MHz and do not affect the quality of experiments in decameter wavelength radio astronomy.

Due to the spatial isotropy of the radio source being searched for, such low-frequency negative phenomena as refraction, scattering, and flicker in the ionosphere, interplanetary and interstellar environments are irrelevant. Ionospheric absorption at frequencies > 8~MHz can be neglected.

Thus, as described in subsection~B, at the observing stage, a high time-dependent resolution of $\Delta t_{p} < 10$ ms is required, 
making possible cleaning from impulse noise (using, in particular, machine learning methods). At the secondary processing stage, averaging over a the required value of $\Delta t \approx 4 \cdot 10^{6}$ s is applied. 

Eliminating radio interference of natural origin caused by Solar and Jupiter bursts with a rich time-frequency structure and maximum intensity at decameter wavelengths also requires a rather high time-frequency resolution at the observing stage. As in the case of other types of RFI, appropriate filtering at the secondary processing stage is possible through machine learning.

\subsection {Recombination lines}

The value $\Delta T_{LR}$ holds a special place among all other interfering effects captured by the radio telescope from the environment. First of all, it is the only factor with a negative polarity, like the hydrogen line $\Delta T_{L}$. If we take into account that the value of $\Delta T_{LR}$ is three orders of magnitude larger than $\Delta T_{L}$, it seems that the recombination lines present everywhere in the Galaxy are very confusing in the process of their reliable separation from the extragalactic effects. However, it is no coincidence that section III describes the details of the physics of the formation of these unusual lines and all their parameters. Consequently, these lines become the most deterministic and stable interfering component against the background of all others, which are extremely non-stationary and sporadic. Thus, the small width of the lines described above $\Delta \nu_{DPR}= 1-10\,\, \mbox{kHz} \ll \Delta \nu_{L} \approx 25\,\, \mbox{MHz}$, their frequency $\nu_{RL}$ and distance between them $\Delta \nu_{RL} \approx 100$ kHz ( the HI line profiles with a width of  $\Delta \nu_{L} \approx 25$~MHz contain approximately 200 recombination lines), the intensity of these lines depending on the physical conditions in the medium and the line distribution in the Galaxy allow us to hope for effective identification and elimination of recombination radio lines from broadband spectra containing cosmological lines of neutral hydrogen within the domain of decameter radio waves.

Next let us consider three more types of parasitic spectral effects (serious interference factors), which, unlike the previous ones, occur in the radio telescope itself and have been poorly described in previous studies.

\subsection{$\Delta G$ hardware noise}

Each radio telescope consists of a large number of radio equipment devices and elements, e. g., antennas, filters, amplifiers, phase shifters, transmission lines, signal splitters, attenuators, and recorders. All of them have frequency-dependent transmission characteristics, which are caused by deterministic active and reactive elements, as well as unpredictable parasitic capacitances, inductances, and resistors \cite{18}.  Consequently, uneven frequency transmission functions are generated with characteristic oscillations with a period of units to tens of MHz and an amplitude of $\Delta G = T_{B}\Delta K$, which coincides with the width of the expected HI line. An additional danger is posed by the fact that these frequency oscillations occur around a certain average power spectrum $G(\nu) \sim T_{B}(\nu)$ and, therefore, may feature negative polarity. Careful adjustment of the elements allows minimizing these fluctuations, but it is impossible to reduce them to less than 1\% of the average. Thus, this parasitic spectral effect can exceed the expected intensity of the HI line by four orders of magnitude, which requires the use of special observation methods (see below).

\subsection{$\Delta P$ hardware noise}

The interference effect, which is labeled in Fig.~\ref{fig14} as $\Delta P \propto T_{B} \Delta K_{2}$  looks similar to the previous $\Delta G$,  but is based on a different physics of origin and is dangerous, too. From the theory of long signal transmission lines, it is known that the power of a reflected noise signal with a uniform spectrum from the loads connected by the long line is
\begin{equation}
P_{bi\nu} (f) =  r^{2} \cos (4\pi L f/V_p)\,, \nonumber
\end{equation}
where $r$  is the coefficient of reflection from loads $(r= 0-1)$, $L$ is the line length, $V_p$ is the phase velocity of radio wave propagation in the transmission line (radio frequency cable), and $f$ is the spectrum frequency.

As is shown, for typical radio engineering parameters of UTR-2, URAN, and GURT, namely, $r \approx 0.1,\,\, L \approx 1-100$ m; $V_p = 200 000$ km/s, it follows that the amplitude of oscillations along the frequency is $r^{2} = 0.01\,\, (1 \%)$ and the period of parasitic oscillations for typical lengths of connecting cables in the radio telescope structure from 1 m to 100 m falls within the range from 1~MHz to 100~MHz. Thus, this interference effect can also coincide with the line width of $\Delta \nu_{L} \approx 25\,\, \mbox{MHz}$,
and have both positive and negative polarity, and is quite powerful, i.\,e. four orders of magnitude greater than the intensity of $\Delta T_{L}$. 

\subsection{$\Delta Q$ hardware noise}

The eventual interference effect of $\Delta Q \propto T_{B}\Delta K_{3}$  was recently discovered by Ukrainian radio astronomers \cite{19}.  Its manifestation is also similar to the previous two effects (power fluctuations of a uniform noise spectrum as a function of frequency). However, in this case, the effect is associated with the placement of antenna elements at a certain height above the ground, especially when the latter is two- or multi-layered. As calculations \cite{19} show, there are multipolar oscillations relative to the mean value with an amplitude of up to 10\% and a period of units to tens of MHz. As in the previous cases, this can be very similar to the spectral effect sought as the cosmological HI line, is dangerous and requires special methods of noise-free observations. 

In every radio astronomy experiment, one has to deal with a classic observational problem: detecting a faint effect (spectral line, signal from a continuous radio source, pulsed or sporadic signal) against a background of powerful noise radiation with a broad spectrum. This can be the intrinsic noise of the radio astronomy receiver, or the brightness temperature of the galactic background, as at decameter wavelengths. To eliminate this powerful ``pedestal'', the method of comparison with the noise spectrum of a reference source, which is subtracted from the spectrum received by the radio telescope, is traditionally used. Different approaches to this method are called modulation or compensation \cite{7}. Fig.~\ref{fig14} shows such a reference source, i.\,e. noise generator NG1, the power of which is regulated by attenuator AT1. It is desirable that the power and spectrum of such a generator $T_{BN}(\nu)$ be as close as possible to the background parameters $T_{B}(\nu)$, which is not easy to achieve in practice. Sometimes a numerical approximation of the broadband spectrum is used (it is assumed that all interference effects have already been eliminated) at the secondary processing stage, but it is required to make sure that the dynamic range of space signal registration and the accuracy of the approximation by a polynomial of a certain degree are adequate.

\begin{figure}[htb]
\includegraphics[width=0.48\textwidth]{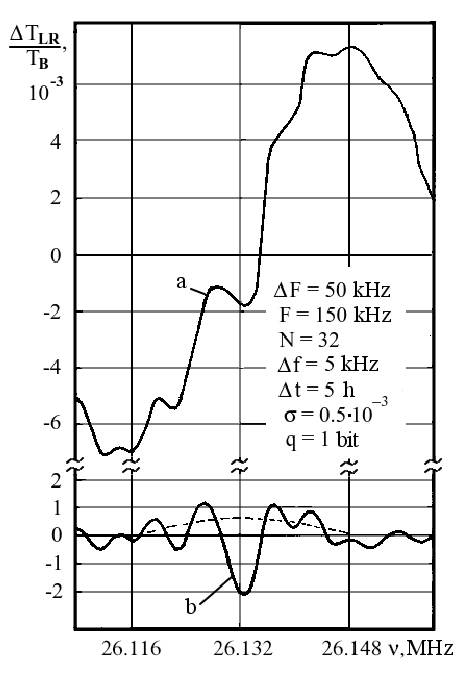} 
\caption{Detection of interstellar recombination radio lines at decameter waves.}
\label{fig15}
\end{figure}

The method of comparison with the spectrum of a reference noise generator was used in experiments to search for and study recombination lines at decameter wavelengths more than 40 years ago \cite{20}.  Fig.~\ref{fig15} shows one of the first spectra (а) measured by UTR-2 in the summer of 1978. It is the difference between the spectrum of the cosmic radio emission of the Cas~A radio source $S_{A}(\nu)$ and the spectrum measured separately without a radio telescope, but using the noise generator NG1 with the spectrum of $S_{\mbox{\tiny NG}} (\nu)$ connected to the spectroanalyzer, similarly to the one shown in Fig.~\ref{fig14}. As can be seen, the ``pedestal'' is absent, but the spectrum is distorted (not flat, the baseline is not equal to 0), primarily due to the above-mentioned significant gradient of the frequency-dependent galactic background. Looking at this measured spectrum, it is almost impossible to assert the detection of a spectral line; an adequate processing methodology must be used. In \cite{20}, a new procedure for correcting the baseline (``zero'') line was proposed, which is simpler, unambiguous, and more efficient than the traditional polynomial approximation, when there is always a problem with choosing the degree of the polynomial and the number of approximation points.

Let’s recall that for decameter radio spectroscopy, the first radio astronomical digital correlation spectroanalyzer on integrated circuits in the former USSR was developed and implemented
at the UTR-2 about 50 years ago \cite{21}. This high-performance device measured the autocorrelation function of the signal under study using the following algorithm
\begin{equation}
R(nT)= \Sigma^{M}_{i=1} x(iT) x(iT + nT)\,,
\label{RnT} \nonumber
\end{equation}
where $x(iT)$ is the input signal sampled in terms of time and amplitude, $T$ is the sampling period (frequency $F= 1/T$), $i$ is the input signal reference number, $n$ is the reference number of the autocorrelation function, $M$ is the total number of samples of the input signal, $MT$ is the total accumulation time in the correlation meter, and $N=n_{max}$ is the total number of channels in the correlation meter.

The comparison method was implemented directly in the correlation meter by measuring the differential autocorrelation function as the difference between the functions from the antenna and the noise generator
\begin{equation}
R(nT) = R_{A} (nT) - R_{\mbox{\tiny NG}}(nT)\,.
\label{RnT2} \nonumber
\end{equation}

At the secondary processing stage, the diffraction spectrum was calculated by direct discrete Fourier transform 
\begin{eqnarray}
S(\nu)&=&S_{A}(\nu)- S_{\mbox{{\tiny NG}}}(\nu)\nonumber\\
&=&2T \Sigma^{N}_{n=1} R(nT) \cos (2\pi \nu nT)\,,
\label{Snu}
\end{eqnarray}
as shown in Fig.~\ref{fig15} (a).

It is clear that these ``smooth'' distortions of the broadband spectrum are generated by the initial samples of the autocorrelation function, while all samples, including $n_{max}=N$, are responsible for the narrow-band spectral line. 

Thus, it was shown for the first time in \cite{20} that ``cutting off'' the initial section of the autocorrelation function $R(nT)$ by the number of $n=N_{K}$ on condition that $N_{K} \ll N$ before the Fourier transform completely eliminates broadband spectral distortions without affecting narrow-band spectral details. This is demonstrated in Fig.~\ref{fig15} (б), when the corrected spectrum is obtained by an extremely simple transformation, virtually indistinguishable from (\ref{Snu})
\begin{equation}
S(\nu)=2 T  \Sigma^{N}_{n=N_{K}} R (n T) \cos (2 \pi \nu n T)\,, 
\label{Snu2} \nonumber
\end{equation}
where $N_{K} = 6$.

Since the developed automated correlation meter \cite{21} measures a normalized function $(R(0)=1)$,  the differential spectrum becomes centered, i.\,e.
\begin{equation}
\int^{\infty}_{0} S(\nu) d \nu = 0\,.\nonumber
\end{equation}

As can be seen from Fig.~\ref{fig15} (b), a small shift of the observed spectral line is 
easily centered with respect to the line maximum, as shown
by the dashed line.

Using this transformation with a cutoff of section $R(n T)$ $n = N_{K} = 3-6$ long, many thousands of measured spectra were quickly processed, which showed the high efficiency of the procedure and fully proved the reliability of detecting a new astrophysical phenomenon, i.\,e. interstellar recombination absorption lines of record-highly excited atoms at extremely low radio frequencies.

The above historical scientific and technical results show not only the possibility of effective correction of the ``zero'' line in the spectra. They also demonstrate the correct choice of the parameters of the spectroscopic search experiment in accordance with the expected parameters of interstellar low-frequency spectral lines (validation of the choice of parameters for the search for cosmological HI lines): 
\begin{itemize}
\item the line frequency  $\nu_{L}$ = 26.13 MHz (decameter waves; neutral hydrogen line, or recombination line;
\item the absolute intensity  $\Delta T_{L} \sim 30$ K;
\item the background temperature $T_{B} \sim$ 30 000 K;
\item the relative intensity $\Delta T_{L}/T_{D} \sim 10^{-3}$;
\item the line width   $\Delta \nu_{L} = 10$ kHz ($\sim 100$ km/s);
\item no polarization;
\item the source angular size $\Theta_{S} = 0.5^{\circ} - 10^{\circ}$.
\end{itemize}

Then the parameters of the experiment, which were formulated about 50 years ago, are as follows (in terms presented at the end of section ІІІ).
\begin{itemize}
\item The analysis bandwidth     $\Delta F = (2-10) \Delta \nu_{L} = 20-100$~kHz.
\item Frequency resolution   $\Delta f = \Delta \nu_{L} / (1-10) = 1-10$~kHz.
\item The number of frequency channels   $N = \Delta F/\Delta f_{min} = 20 - 100$.
\item Sampling frequency  $F \geqslant 2 \Delta F = 40- 200$ kHz.
\item Sampling quantization   $q = 1-16$ bit.
\item Spectral resolution adjusted to the width of radio interference or recombination lines$\Delta f_{RFI} = \Delta f = 1-10$ kHz.
\item The number of channels  $N_{RFI} = N = 20-100$.
\item The space-based resolution   $\Theta_{A} = \Theta_{S}= 0.5^{\circ}-10^{\circ}$.
\item The relative sensitivity        $\Delta T_{e}/T_{D} \approx 10^{-3}$.
\item The signal to noise ratio $S/N \geqslant 10$.
\item The fluctuation level at spectra\\  $\sigma \leqslant (\Delta T_{L}/T_{B})/(S/N) = 10^{-4}$.
\end{itemize}

According to (7), the required integration time for the $\Delta f$ resolution from 1 kHz to 10 kHz is estimated as follows:
\begin{equation}
\Delta  t = 1/(\Delta f \sigma^{2}) = 10^{4} - 10^{5} \mbox{sec}\approx 3- 30\,\, \mbox{hours}\,.\nonumber
\end{equation}
The selected
parameters of the experiment are shown in Fig.~\ref{fig15}. Implemented at the UTR-2 radio telescope, they enable detecting a very faint radio spectroscopic effect successfully. This is also confirmed by much more sensitive experiments \cite{13}, shown in Figs. \ref{fig12}, \ref{fig13}. With an analysis bandwidth of $\Delta F \approx 1.5$ MHz and an effective accumulation time of $\Delta t = 5000$ hours, a record-breaking sensitivity of $\sigma \approx 10^{-5}$ was achieved. Obviously, when the data with a resolution of $\Delta f \approx 1$ kHz is ``smoothed'' to $\Delta f \approx 25$ MHz, the sensitivity will become close to $\sigma = 10^{-6} - 10^{-7}$, which is required to find the HI line. This proves the good prospects for solving many astrophysical problems related to radio spectroscopy of cosmic radiation at decameter wavelengths.

Still, there is a fundamental difference between the above experiments on recombination lines and the new task of searching for the cosmological HI line at decameter wavelengths.

First of all, the four key parameters of the experiment, i.\,e. the width of the $\Delta \nu_{L}$ line, the analysis band $\Delta F$, the number of channels $N$, and the relative intensity of $\Delta T_{L}/T_{B}$ differ by a factor of three to four orders of magnitude. As can be seen, when studying narrow recombination lines within the accordingly selected analysis band, there is virtually no interference from $\Delta I_{RFI}$ (one can find RFI-free ``windows''), the $\Delta I_{ATM}, \Delta G, \Delta P, and \Delta Q$ factors are much wider than $\Delta \nu_{L}$ and even $\Delta F$. The latter only form a certain additional gradient in the frequency band of $\sim 50$~kHz, which is eliminated effectively, as shown above. The temperature $T_{B}$ is not detrimental for observations of recombination lines either, since the absolute intensity of the latter depends on $T_{B}$: $\Delta T_{LR} = T_{B}\sigma_{LR}$. Thus, the 
searching for the cosmological HI line proposed in this paper is much more complex and requires special methods for eliminating numerous interfering factors described at the beginning of section III.

It should be noted that the method of comparison with the NG1 signal shown above is quite effective, but imperfect. As can be seen from Fig.~\ref{fig14}, the generator signal does not cover a significant part of the radio telescope elements and systems -- antennas, amplifiers, a part of the transmission lines, etc. In the spectrum from NG1, there are virtually no interfering factors, either external or internal, that would be subtracted from the interference in the spectrum received by the antenna during the comparison process. The comparison method with diagrammatic modulation can significantly improve the situation. In this case, the main beam of the radio telescope (beam pattern) is directed at the radio source under study, which has a rather compact size $\Theta_{S}$, with the corresponding spectrum $S'(\nu)$ recorded. Another beam with the same width $\Theta_{A}$ is pointed (sequentially or simultaneously) at a neighboring part of the sky at a distance of $\Delta \Theta > \Theta_{S} (\Theta_{A} < \Delta \Theta)$, but as close to the source as possible so that the telescope parameters do not change. Besides, a comparative spectrum $S'' (\nu)$ is recorded, too. It is very important that this spectrum has all the same external and internal interfering spectral features. Then, when obtaining the differential spectrum $S(\nu)=S'(\nu) - S''(\nu)$, we only have the positive effect of $\Delta T_{L}$, and all others are compensated. Unfortunately, this method is valid for compact radio sources only. In our case of an isotropic radio source with an HI line, the $\Delta T_{L}$ effect will also be destroyed when obtaining a differential spectrum.

In this paper, we propose another, more efficient method of comparing the spectra, namely, we use spectra measured in the directions of the Galactic background with temperatures $T'_{В}$ and $T''_{B}$, that differ by a factor of about two (see Fig.~\ref{fig11}). Both spectra contain all the interfering factors (in slightly different values) and a positive spectral feature. This scheme can be presented analytically. For simplicity, let us assume that all interference with the narrow-band structures $\Delta T_{LR}, \Delta I_{RFI}, \Delta I_{ATM}$ have already been eliminated. Also please, note that the positive effect of $\Delta T_{L}$ (as well as the narrow-band interference factors) is additive to $T_{B}$, while interferences $\Delta G, \Delta P, \Delta Q$ are multiplicative with respect to $T_{B}$. Consequently, we can record the two corresponding spectra as follows
\begin{eqnarray}
\hskip-0.7cm S'(\nu)=T'_{B} + \Delta T_{L} + T'_{B} \Delta K + T'_{B} \Delta K_{2} +T'_{B} \Delta K_{3}\,,\label{S'nu} \nonumber\\ 
\hskip-0.3cm S''(\nu)=T''_{B} + \Delta T_{L} + T''_{B} \Delta K_{1} + T''_{B} \Delta K_{2} + T''_{B} \Delta K_{3}\,.\label{S''nu} \nonumber
\end{eqnarray}
For simplicity, let’s assume that $T''_{B} \approx 2 T'_{B}$, then
\begin{eqnarray}
S'' (\nu) \approx 2 T'_{B} &+& \Delta T_{L} +2 T'_{B} \Delta K_{1} \nonumber\\
&+& 2 T'_{B} \Delta K_{2} + 2 T'_{B} \Delta K_{3}\,.\nonumber 
\end{eqnarray}

To equalize the total powers of both spectra (as in the case of equalizing the powers of the spectra from the antenna and the noise generator), we double the value of $S'(\nu)$ and obtain the differential spectrum as follows
\begin{equation}
S(\nu) \approx 2S' (\nu) - S'' (\nu) \approx 2 \Delta T_{L} - \Delta T_{L} = \Delta T_{L}\,. \label{Snu3} \nonumber
\end{equation}

Thus, the change in the temperature of the Galactic background (approximately by a factor of 2) due to the diurnal rotation of the Earth with a fixed radiation pattern, the same change in the interfering factors proportional to $T_{B}$, and the invariability of the intensity of $\Delta T_{L}$ allow distinguishing the latter from the Galactic background itself (a large ``pedestal'') and from the dangerous broadband (close to the HI line) interfering factors. Definitely, the quality of these comparisons of $S'(\nu)$ and $S''(\nu)$ depends on the coincidence of the corresponding spectra, one of which tends to the Galactic plane, and the other to the high-latitude regions of the Galaxy. 

The receiver-recorder of cosmic radio signals plays a very important role in the signal path of each radio telescope (Fig.~\ref{fig14}). Its construction depends on the characteristics of the signals under study. In our case, we need a spectrum analyzer, i.\,e., a device measuring the power of the input signal within rather narrow bands distributed sequentially over the widest possible frequency band. Several constantly progressing generations of correlation spectrum analyzers were developed for the UTR-2 radio telescope \cite{21,22,13}  with a frequency bandwidth increasing from $\sim100$ kHz to $\sim30$ MHz, and the number of channels from 32, 128 to 4096 (the first device was briefly mentioned above). This approach proved to be quite effective. Measuring the autocorrelation function of a signal before its Fourier transform is extremely useful because it allows correcting the ``zero'' line, quickly detecting narrow-band radio interference, and ensures the maximum signal accumulation with minimal computer memory. Single-bit quantization in such analyzers simplifies the design as much as possible, ensuring stability and speed. However, despite the advantages, single-bit quantization reduces the noise immunity of broadband measurements ($\Delta F \sim 50$~MHz) in the presence of narrow-band intense radio interference, which is inherent in our proposed research. There are two obvious solutions to this problem: when building a new correlation spectrum analyzer, it is required to use an analog-to-digital converter with at least 16 bits, as well as special chip-based and computer equipment that is now available. The second approach is to measure the spectrum by digital direct fast Fourier transform of the real-time input signal in accordance with the following ratio
\begin{eqnarray}
G(\nu)&=&\frac{1}{M}\Bigl\{\left[\Sigma^{M}_{i=0} x(i T) \cos (2\pi \nu i T)\right]^{2} \nonumber\\ 
&+&\left[\Sigma^{M}_{i=0} x(i T) \sin (2 \pi \nu i T)]^{2}\right]^2\Bigr\}\,. \label{Gnu} \nonumber
\end{eqnarray}

Such new-generation spectrum analyzers have already been developed and used successfully at the UTR-2, URAN, and GURT radio telescopes \cite{4}. Their important feature is 
extreme versatility and possibility measuring all parameters of the cosmic radio signals, i.\,e. power, spatial, spectral, temporal, polarization, cross-correlation, with the highest sensitivity and noise immunity.
Both direct measurements of spectra and calculations of autocorrelation and cross-correlation functions are possible (these spectrum analyzers are proposed for the search for cosmological HI lines, at least at the first stages of observations).

A prospective algorithm for these observations is as follows.
\begin{enumerate}
\item The antenna system of the radio telescope (one dipole or subarray of the GURT with a bandwidth of 8--80 MHz) shall be phased in the zenith direction (the height above the horizon $h$ is $90^{\circ}$) and the equatorial coordinate (direct declination) is $\delta=50^{\circ}$.
\item  During $\sim$ 12 hours (near the maximum of the galactic background), the power spectra at the antenna output shall be measured with a given pre-accumulation interval of $\Delta t_{n} = 0.1-1$ sec (this corresponds to the time-based resolution at a frequency-based resolution of $\Delta f = 4$~kHz). This sequence of spectra $S''(\nu)$ (source data) shall be saved in the computer memory.
\item  The same procedure shall be carried out within the next 12 hours near the background minimum, and the sequence of spectra $S'(\nu)$ shall also be saved in the computer's memory.
\item  After diurnal observations, it is advisable to conduct secondary operational processing, i.\,e. filtering out narrow-band and impulse interference on all partial spectra, smoothing the spectra to resolutions, for example, $\Delta f = 2.5$ MHz and $\Delta f = 25$~MHz, and time averaging to $\Delta t = 12$ hours for $S'(\nu)$ and $S''(\nu)$ spectra. These actions significantly compress the information that should be stored on hard disks. At the same stage, the differential spectra of $S(\nu) = 2 S'(\nu) - S''(\nu)$ may be calculated and accumulated on the same disks.
\item Steps 1--4 shall be carried out diurnally for six months to cover all positions of the Milky Way in the sky (e. g., from winter to summer) with a clear time account for the diurnal shift in the Milky Way position by 3 min 57 sec per day. All spectra recorded during six months shall also be saved on disks for the final processing and data analysis. The latter includes, but not limited to, averaging the data over time to the maximum accumulation time of $\Delta t$.
 \end{enumerate}

The key criterion for the accuracy of the experimental methodology is the assessment of the fluctuation level in the spectra and its compliance with formula $\sigma = 1/ \sqrt{\Delta f \Delta t}$, taking into account the frequency resolution and accumulation time. The effect of eliminating all types of interference shall also be evaluated by a clear drop in the fluctuation level in accordance with the $\sigma \sim 1/\sqrt{\Delta t}$ law.

\section{Some prospects for improving the methodology of searching for the cosmological HI line at decameter wavelengths}

As follows from the previous sections, the search for 
the cosmological phenomenon is an extremely difficult observational radio astronomy problem. It requires 
record-breaking parameters of equipment, and methods that have never been implemented before in decameter-wavelength radio astronomy. The tools and methods described above allow us to hope for positive results. However, let us point out at some additional opportunities for improving the chances of successful search for HI cosmological lines.

\subsection{Multi-antenna observations}

The Galactic non-thermal radio emission is a broadband $\delta$-correlated noise distributed over the sky. When observations are made using independent antennas spaced at a distance of $d> \lambda/2$ (at decameter-meter wavelengths $d>50\,\, \mbox{m} - 0.5\,\, \mbox{m}$), the signals from the background at the antenna outputs will be decorrelated within the corresponding frequency range of 3 MHz to 300 MHz. Thus, these signals can be averaged, which will lead to a decrease in fluctuations by a factor of $\sqrt{R}$ ($R$ is the number of radio telescopes or dipoles), which is equivalent to a decrease in the required accumulation time by a factor of $R$. Since a single antenna element is a fairly simple design, they can be easily used in a quite large number -- $R \gg 1$. The situation is complicated by the fact that the number of spectrum analyzers must be the same. The GURT subarray is more complex than the element, but there are more than 10 of them, and subject to availability of the appropriate number of recorders, the multi-antenna methodology is also possible.

\subsection{Expansion of the frequency range of UTR-2 and GURT antennas}

The antenna elements and subarrays included in the UTR-2 and URAN radio telescopes (the total number of elements is about 4000) are traditionally used within the 8--32 MHz band, which is not sufficient to search for HI lines. However, there is a reason to believe that the actual permitted frequency band of the telescopes is wider,  5--50~MHz. This needs further investigation. If the conclusion is positive, multi-antenna studies will be possible on a large number of antennas of these radio telescopes, too.

\subsection{Dual-beam observations}

The GURT, UTR-2, and URAN subarrays have relatively narrow radiation patterns, $\Theta_{A} \sim 30^{\circ}$. If we direct the pattern of one subarray to the point with coordinates $\delta=50^{\circ}, \alpha = 0^{n}$, and the other to $\delta= 50^{\circ}, \alpha = 12^{h}$ (however, symmetrically to the polar star and the Earth's surface) from the measurements of the corresponding spectra, as shown above, we obtain the similarity of $S'(\nu)$ and $S''(\nu)$ spectra in the ``counter-phase'' and the identity of the radio interference, which are further reduced when obtaining differential spectra, and the observation time is halved.

\subsection{Modeling experiments for testing the methodology}

The possibility of achieving record-breaking radio spectroscopic sensitivity can be tested using an instrumental model. In this case, the experimental setup and methodology remain unchanged, but instead of the antenna, the devices simulating the radio astronomy conditions are connected. These are noise generator NG2, the power of which is regulated by attenuator AT2 to the $T_{B}$ level with a spectrum of 8--80 MHz, and NG3, the signal of which is added to the signal of NG2 through a filter with a bandwidth of $\Delta \nu_{L}= 25$ MHz, but with a strong attenuation of AT3 to ensure the ratio of $\Delta T_{L}/T_{B} \approx 10^{-6}$. This should confirm the required accumulation time of $\Delta t$, and the absence of a certain amount of internal interference.

\subsection{Variational method of multiple comparison}

To search for new faint radio astronomy effects, it is extremely important to use different independent experimental facilities located in different geographical locations and with slightly different experimental parameters, followed by comparison of the results. Even some changes in parameters during long-term experiments will be desirable, such as the type of dipoles, length of transmission lines, amplifiers, spectrum analyzers with different analysis bands and resolutions. When comparing the results, the effect you are looking for should be repeatable in any configuration of devices and parameters. \\

\subsection{Lunar missions}

Undoubtedly, the decisive step to fundamentally improve the conditions for searching the cosmological HI lines and other faint astrophysical effects at low frequencies is to install decameter-hectometer-wavelength radio telescopes on the far side of the Moon. This will create a unique opportunity to avoid numerous terrestrial RFI signals, and the negative influence of the ionosphere.
There is no doubt that Lunar projects of this kind will be implemented over the next 10--20 years.

\section*{Conclusions} 

The radio spectroscopic experiment proposed and described herein is extremely complex, time-consuming, but not hopeless. Theoretical estimates of the possibility of the existence of neutral hydrogen lines redshifted into the domain of decameter wavelengths and their physical parameters justify prospective observational efforts. 
The instrumental requirements for such the observational program 
are based on 50 years of experience in radio astronomy research at decameter wavelengths, including radio astrospectroscopy. Such the experimental program appears to be feasible in the coming years.

\section*{Acknowledgements} The work was supported by the projects of the National Academy of Sciences of Ukraine (Horizon, Interstellar, Radio Telescope) and departmental research topics of the Department of Physics and Astronomy of the National Academy of Sciences of Ukraine, as well as the project of the Ministry of Education and Science of Ukraine Modeling the Luminosity of the Large-Scale Structure Elements of the Early Universe and Galactic Supernovae Remnants and Observations of Variable Stars (state registration number 0122U001834). Vyacheslav Zakharenko is grateful for the financial support of the Europlanet 2024 RI project funded by the European Union's Horizon 2020 Research and Innovation Program (grant agreement No. 871149). The authors are grateful to their European colleagues for their participation in the preparation of the UPSCALE program proposals.

\newpage


\begin{thebibliography}{3}
\bibitem {Penzias1965} A. A. Penzias, R. W. Wilson, Astrophys. J. {\bf 142}, 419 (1965); https://doi.org/10.1086/148307
\bibitem {Planck2020a} Planck Collaboration: N. Aghanim et al.,  Astron. Astrophys. {\bf 641}, A1 (2020); https://doi.org/10.1051/0004-6361/201833880
\bibitem {Planck2020b} Planck Collaboration: N. Aghanim et al., Astron. Astrophys.,  {\bf 641}, A6 (2020); https://doi.org/10.1051/0004-6361/201833910
\bibitem {Robertson2023} B. E. Robertson et al., Nature Astronomy {\bf 7}, 611 (2023); https://doi.org/10.1038/s41550-023-01921-1 
\bibitem {Curtis-Lake2023}  E. Curtis-Lake, Four metal-poor galaxies spectroscopically confirmed beyond redshift ten, submitted to Nature Astronomy.
\bibitem {Verde2019} L. Verde, T. Treu, A.G. Riess, Nature Astronomy {\bf 3}, 891 (2019); https://doi.org/10.1038/s41550-019-0902-0
\bibitem {Riess2022} A. G. Riess et al., Astrophys. J. Lett. {\bf 934}, id. 7 (2022); https://doi.org/10.3847/2041-8213/ac5c5b
\bibitem {Bromm2011} V. Bromm, N. Yoshida, Ann. Rev. Astron. Astrophys. {\bf 49}, 373 (2011); https://doi.org/10.1146/annurev-astro-081710-102608
\bibitem {Pritchard2012} J. R. Pritchard, A. Loeb, Rep. Prog. Phys. {\bf 75}, id. 086901 (2012); https://doi.org/10.1088/0034-4885/75/8/086901
\bibitem {Bowman2018} J. D. Bowman et al., Nature {\bf 555}, 67 (2018); https://doi.org/10.1038/nature25792 
\bibitem {Singh2022} S. Singh et al., Nature Astronomy, {\bf 6}, 607 (2022); https://doi.org/10.1038/s41550-022-01610-5 
 \bibitem {Novosyadlyj2020} B. Novosyadlyj, Yu. Kulinich, V. Shulga, W. Han, Astrophys. J. {\bf 888}, id. 27 (2020); https://doi.org/10.3847/1538-4357/ab530f
\bibitem {Kulinich2020} Yu. Kulinich, B. Novosyadlyj, V. Shulga, W. Han, Phys. Rev. D {\bf 101}, id.083519 (2020); https://doi.org/10.1103/PhysRevD.101.083519
\bibitem {Novosyadlyj2022} B. Novosyadlyj, Yu. Kulinich, B. Melekh, V. Shulga, Astron. Astrophys. {\bf 663}, A120 (2022); https://doi.org/10.1051/0004-6361/202243238
\bibitem {Novosyadlyj2020a} B. Novosyadlyj, V. Shulga, Yu. Kulinich, W. Han, Physics of the Dark Universe {\bf 27}, 100422 (2020); https://doi.org/10.1016/j.dark.2019.100422
\bibitem {Novosyadlyj2023} B. Novosyadlyj, Yu. Kulinich, G. Milinevsky, V. Shulga, Mon. Not. Roy. Astron. Soc. {\bf 526}, 2724 (2023); https://doi.org/10.1093/mnras/stad2927 
\bibitem {Tauscher2018} K. Tauscher, D. Rapetti, J. O. Burns, E. Switzer, Astrophys. J. {\bf 853}, id. 187 (2018); https://doi.org/10.3847/1538-4357/aaa41f
\bibitem {Rapetti2020} D. Rapetti, K. Tauscher, J. Mirocha and J. O. Burns, Astrophys. J. {\bf 897}, id. 174 (2020); https://doi.org/10.3847/1538-4357/ab9b29 
\bibitem {Tauscher2020} K. Tauscher, D. Rapetti, J. O. Burns, Astrophys. J. {\bf 897}, id. 175 (2020); https://doi.org/10.3847/1538-4357/ab9b2a
\bibitem {Tauscher2021} K. Tauscher et al., Astrophys. J. {\bf 915}, id. 66 (2021); https://doi.org/10.3847/1538-4357/ac00af
\bibitem {Novosyadlyj2023a} B. Novosyadlyj, Y. Kulinich, O. Konovalenko, Journal of Physical Studies {\bf 28}(1), 1901 (2024);
https://doi.org/10.30970/jps.28.1901
\bibitem {1} S. Ya. Braude, Antennas {\bf 26}, 3 (Svyaz’ Publ., Moscow, 1978).
\bibitem {2} O. O. Konovalenko et al., Radio Phys. Radio Astron. {\bf 26}, 5 (2021); https://doi.org/10.15407/Rpra26.01.005
\bibitem {3} A. Konovalenko et al., Exp. Astron. {\bf 42}, 11 (2016);  https://doi.org/10.1007/ S10686-016-9498-X
\bibitem {4} V. Zakharenko et al., J. Astron. Instrum. {\bf 5}, 1641010 (2016); https://doi.org/10.1142/S2251171716410105
\bibitem {5} A. А. Konovalenko et al., Radio Phys. Radio Astron. {\bf 21}, 83 (2016); https://doi.org/10.15407/rpra21.02.083
\bibitem {6} O. O. Konovalenko  et al., Radio Phys. Radio Astron. {\bf 24}, 3 (2019); https://doi.org/10.15407/rpra24.01.003
\bibitem {7} J. D. Kraus, {\it Radio Astronomy} (McGraw-Hill Inc., US, 1966).
\bibitem {8} P. L. Tokarsky, A. A. Konovalenko, S. N. Yerin, IEEE Trans. Antennas Propag. {\bf 65}, 4636 (2017); https://doi.org/10.1109/tap.2017.2730238
\bibitem {9} P. L. Tokarsky, A. A. Konovalenko, S. N. Yerin, I. N. Bubnov, IEEE Trans. Antennas Propag. {\bf 67}, 7312 (2019); https://doi.org/10.1109/tap.2019.2929322
\bibitem {26} М. А. Sidorchuk et al., Radio Phys. Radio Astron. {\bf 26}, 287 (2021); https://doi.org/10.15407/rpra26.04.287
\bibitem {10} V. V. Krymkin, Radiophys. Quantum electron. {\bf 14}, 161 (1971); https://doi.org/10.1007/bf01031395
\bibitem {11} A. A. Konovalenko, L. G. Sodin,  Nature {\bf 294}, 135 (1981); https://doi.org/10.1038/294135a0
\bibitem {12} A. A. Konovalenko, S. V. Stepkin, In: {\it Radio astronomy from Karl Jansky to microjanski}, (Eds.) L. I. Gurvits, S. Frey, S. Rawlings, Vol. 15, pp. 271-295 (EAS publ., Budapest, 2005);  https://doi.org/10.1051/eas:2005158
\bibitem {13} S. V. Stepkin, A. A. Konovalenko, N. G. Kantharia, N. Udaya Shankar, Mon. Not. R. Astron. Soc. {\bf 374}, 852 (2007); https://doi.org/10.1111/j.1365-2966.2006.11190.x
\bibitem {14} A. A. Konovalenko, In: {\it Radio recombination lines: 25 years of investigation}, (Eds.) M. A. Gordon, R. L. Sorochenko, Proceedings of IAU colloq., pp. 175-189 (Springer, Dordrecht, 1990); https://doi.org/10.1007/978-94-009-0625-9\_17
\bibitem {15} S. V. Stepkin, O. O. Konovalenko, Y. V. Vasylkivskyi, D. V. Mukha, Radio Phys. Radio Astron. {\bf 26}, 314 (2021); https://doi.org/10.15407/rpra26.04.314 
\bibitem {16} A. K. Vydula et al., Astron. J. {\bf 167}, id. 2 (2024); https://doi.org/10.3847/1538-3881/ad08ba
\bibitem {17} S. Ya. Braude, A. V. Megn, B. P. Ryabov, N. K. Sharykin, I. N. Zhouck, Astrophys. Space Sci. {\bf 54}, 3 (1978); https://doi.org/10.1007/bf00637902
\bibitem {18} E. P. Abranin, Yu. M. Bruck, V. V. Zakharenko, A. A. Konovalenko, Exp. Astron. {\bf 11}, 85 (2001); https://doi.org/10.1023/a:1011109128284
\bibitem {19} P. L. Tokarsky, A. A. Konovalenko, J. Modelski, IEEE Access {\bf 11}, 75225 (2023); https://doi.org/10.1109/ACCESS.2023.3294694
\bibitem {20} O. O. Konovalenko, {\it Radio spectroscopic studies of the interstellar medium at decameter waves}, PhD Thesis (IRE NASU, Kharkiv, 1982).
\bibitem {21} A. А. Konovalenko, Pribory i tekhnika experimenta {\bf 6}, 123 (1981);
\bibitem {22} S. V. Stepkin, Radio Phys. Radio Astron. {\bf 1}, 255 (1996); http://rpra-journal.org.ua/index.php/ra/article/view/22/278

\end{thebibliography}
\end{document}